\documentclass[aps,twocolumn,superscriptaddress,nobibnotes,a4paper]{revtex4-2}

\usepackage{amssymb}
\usepackage{natbib}
\usepackage{graphicx}
\usepackage{amsmath}
\usepackage[bookmarks = false, backref = false]{hyperref}
\usepackage{bm}
\usepackage[all]{hypcap}
\usepackage{graphicx}
\usepackage{colortbl}
\usepackage{booktabs}
\usepackage{enumerate}
\usepackage{enumitem}
\usepackage{braket}
\usepackage{mathrsfs}
\usepackage{multirow}
\usepackage{upgreek}
\usepackage{textcomp}

\usepackage{soul}
\usepackage{dsfont}
\usepackage{xcolor}
\usepackage{array}

\renewcommand{\andname}{\ignorespaces}
\def\arraystretch{1.2}

\begin{document}

\title{Frequency tunable, cavity-enhanced single erbium quantum emitter\\in the telecom band}

\author{Yong Yu}
\affiliation{Kavli Institute of Nanoscience, Department of Quantum Nanoscience,
Delft University of Technology, 2628CJ Delft, The Netherlands}

\author{Dorian Oser}\thanks{Current address: QphoX B.V., 2628XG Delft, The Netherlands}
\affiliation{QuTech, Delft University of Technology, 2628CJ Delft, The Netherlands}

\author{Gaia Da Prato}
\affiliation{Kavli Institute of Nanoscience, Department of Quantum Nanoscience,
Delft University of Technology, 2628CJ Delft, The Netherlands}

\author{Emanuele Urbinati}
\affiliation{Kavli Institute of Nanoscience, Department of Quantum Nanoscience,
Delft University of Technology, 2628CJ Delft, The Netherlands}

\author{Javier Carrasco \'Avila}
\affiliation{Department of Applied Physics, University of Geneva, 1211 Geneva, Switzerland}
\affiliation{Constructor University Bremen GmbH, 28759 Bremen, Germany}

\author{Yu Zhang}
\affiliation{Kavli Institute of Nanoscience, Department of Quantum Nanoscience,
Delft University of Technology, 2628CJ Delft, The Netherlands}

\author{Patrick Remy}
\affiliation{SIMH Consulting, Rue de Gen\`eve 18, 1225 Ch\^ene-Bourg, Switzerland}

\author{Sara Marzban}\thanks{Current address: MESA+ Institute for Nanotechnology, University of Twente, 7500AE Enschede, The Netherlands}
\affiliation{QuTech, Delft University of Technology, 2628CJ Delft, The Netherlands}

\author{Simon Gr\"oblacher}
\email{s.groeblacher@tudelft.nl}
\affiliation{Kavli Institute of Nanoscience, Department of Quantum Nanoscience,
Delft University of Technology, 2628CJ Delft, The Netherlands}

\author{Wolfgang Tittel}\email{wolfgang.tittel@unige.ch}
\affiliation{QuTech, Delft University of Technology, 2628CJ Delft, The Netherlands}
\affiliation{Department of Applied Physics, University of Geneva, 1211 Geneva, Switzerland}
\affiliation{Constructor University Bremen GmbH, 28759 Bremen, Germany}

\begin{abstract}
Single quantum emitters embedded in solid-state hosts are an ideal platform for realizing quantum information processors and quantum network nodes. Among the currently-investigated candidates, Er$^{3+}$ ions are particularly appealing due to their 1.5~$\upmu$m optical transition in the telecom band as well as their long spin coherence times. However, the long lifetimes of the excited state---generally in excess of 1 ms---along with the inhomogeneous broadening of the optical transition result in significant challenges. Photon emission rates are prohibitively small, and different emitters generally create photons with distinct spectra, thereby preventing multi-photon interference -- a requirement for building large-scale, multi-node quantum networks. Here we solve this challenge by demonstrating for the first time linear Stark tuning of the emission frequency of a single Er$^{3+}$ ion. Our ions are embedded in a lithium niobate crystal and couple evanescently to a silicon nano-photonic crystal cavity that provides an up to 143 increase of the measured decay rate. By applying an electric field along the crystal c-axis, we achieve a Stark tuning greater than the ion's linewidth without changing the single-photon emission statistics of the ion. These results are a key step towards rare earth ion-based quantum networks.
\end{abstract}

\maketitle

Quantum emitters play a vital role in future quantum networks~\cite{kimble2008,wehner2018}. They serve as the crucial light-matter interface, bridging static qubits that handle local quantum information processing with flying qubits, which are responsible for remote information transfer. Among quantum emitter platforms such as trapped neutral atoms~\cite{sangouard2011,reiserer2015}, ions~\cite{duan2010} and quantum dots~\cite{lodahl2017a}, single defects in the solid state~\cite{awschalom2018} have proven to be among the most successful, owing to their ease of use and potential for large-scale integration. Notably, defects such as nitrogen-vacancy and silicon-vacancy centers in diamond~\cite{ruf2021a} have recently been used for seminal demonstrations of multi-node quantum networks~\cite{pompili2021,hermans2022a} and memory-enhanced quantum communication~\cite{bhaskar2020}, respectively. One major downside for these defects is that they operate at visible or close-to-visible optical wavelengths and are therefore subject to significant transmission loss that exceeds 2~dB/km of fiber, which hampers their integration into large-scale quantum networks~\cite{hensen2015}. In addition, their optical coherence time is typically limited to a few tens of nanoseconds, restricting their suitability for ensemble-based quantum memory for light~\cite{askarani2021}.

Trivalent rare-earth ions (REIs) represent another type of solid-state defect~\cite{thiel2011}. Their $4f$ electrons are effectively shielded by filled outer electronic shells, resulting in narrow linewidths---or long coherence times---for $4f-4f$ transitions in the optical band and spin-state transitions in the microwave regime. This makes them ideally suited as ensemble-based optical quantum memories~\cite{lvovsky2009a}. Using the controlled reversible inhomogeneous broadening (CRIB)~\cite{kraus2006} or atomic frequency comb (AFC) protocols~\cite{afzelius2009}, they achieve high efficiency~\cite{hedges2010,sabooni2013,davidson2020}, long storage times~\cite{ortu2022a, ma2021a, askarani2021}, large bandwidth~\cite{saglamyurek2011}, high fidelity~\cite{zhou2012}, and large multi-mode capacity~\cite{yang2018a}. Using individual REIs (instead of ensembles) furthermore allows the creation of true single photons~\cite{kolesov2012, dibos2018,zhong2018,kindem2020,raha2020,yang2023,gritsch2023,ourari2023,deshmukh2023} and promises quantum information processing with qubits encoded into spin states ~\cite{kinos2021a} that can feature coherence times up to a few hours~\cite{zhong2015a}. This large versatility sets rare-earth-doped crystals apart from any other solid-state impurity.

However, using individual REIs---either for sources or for qubits---is challenging because of their long excited-state lifetimes and hence low photon emission rate. The most practical way to overcome this problem is to Purcell-enhance~\cite{reiserer2015} a spectrally isolated ion within the inhomogeneously broadened transition. By fabricating a nanoscale photonic-crystal (PhC) cavity out of the rare-earth-doped crystal~\cite{zhong2018,yang2023,gritsch2023} or by bringing a separately fabricated PhC cavity in close proximity to the REI-doped material~\cite{dibos2018}, ion emission within the cavity mode is strongly enhanced. Using this approach, single-photon sources and non-demolition measurements have been demonstrated~\cite{kindem2020, raha2020, yang2023, gritsch2023, ourari2023}. The enhanced interaction with light also enables the observation of the coupling between a single REI and neighboring nuclear spins~\cite{ruskuc2022,uysal2023}, and, in the future, between neighboring REIs~\cite{kinos2021a} as well as photon-mediated entangling of distant REIs~\cite{barrett2005}.

Unfortunately, the spectral properties of the emitted photons are largely determined by their local environment, leading to distinguishable spectral lines and spectral diffusion~\cite{thiel2011}. While a quantum eraser technique~\cite{zhao2014} allows entangling photons with a frequency differences up to around 80~MHz, it is generally preferable to use indistinguishable photons for this operation, which is crucial in quantum networks. As a consequence, two-photon interference of the Hong-Ou-Mandel type has so far merely been demonstrated between two consecutive photons from the same REI emitter~\cite{ourari2023}, embedded inside a specific host crystal that limits spectral diffusion but, at the same time, also prevents external spectral control.

\begin{figure}[b!]
	\centering
	\includegraphics[width=0.9\columnwidth]{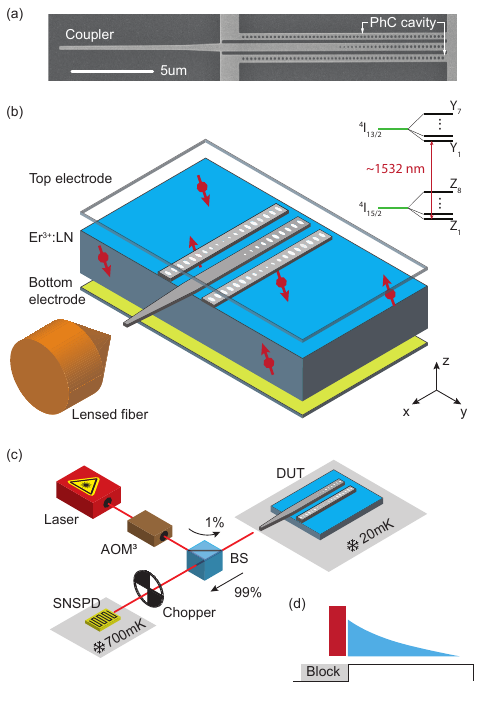}
	 \caption{Experimental setup. (a) A scanning electron microscopic image of the photonic integrated chip (PIC), consisting of two photonic crystal (PhC) cavities and a tapered waveguide in the middle as a common coupler. (b) Schematic of the device. The silicon PIC is placed on an Er:LN crystal and is optically accessed via a lensed fiber. A pair of electrodes above and below the LN crystal allows applying an electric field. The top right inset shows a simplified energy level scheme of Er:LN at no magnetic field. The crystal field splits the $^{2S+1}L_J$ states of the free Er$^{3+}$ ion (green lines) into $J+1/2$ Kramers doublets (black lines). The $Z_1-Y_1$ transition is at around 1532~nm. (c) Optical setup for measuring photo luminescence (PL). Three concatenated acoustic optical modulators (AOMs) are used to create excitation pulses from a continuous wave laser. A 99:1 beamsplitter (BS) routes the excitation pulse to the DUT (only one cavity is shown) and guides the light emitted from the sample to the detection setup. After an optical chopper, which blocks the reflected excitation pulses, a superconducting nanowire single-photon detector (SNSPD) detects the PL signal. (d) Time sequence of the experiment. The exponentially-decaying PL (blue) is collected after a short excitation pulse (red) is sent.}
	\label{fig:setup}
\end{figure}

In this letter, we demonstrate for the first time spectral engineering of a cavity-enhanced single REI emitter. Towards this end we chose Erbium (Er$^{3+}$), which is particularly attractive as its telecom C-band optical transition provides an opportunity to directly access fiber-based networks with minimal loss. More precisely, we use evanescently-coupled silicon nanophotonic cavities to Purcell-enhance and collect the emission from Er$^{3+}$ ions doped into a lithium niobate crystal. We observe an increase in the decay rate of the excited state by up to a factor of 143 (corresponding to a reduction of the lifetime T$_1$ from 1.8~ms to 12.5~$\upmu$s) and measure typical values for auto-correlation coefficients of the emitted light around 0.2, indicating spectrally isolated single ions \cite{gerry2004}. We then control the ion frequency by applying an electric field along the z-axis of the LiNbO$_3$ (LN) crystal. This allows us to demonstrate a frequency tunability greater than the emitter linewidth (measured during around 1~min) without compromising the single-quantum character of the emitter. This promises making different emitters spectrally indistinguishable and countering spectral diffusion through feedback. Our findings directly complement those reported in~\cite{yang2023}, in which an electric field was used to tune the cavity resonance but not that of the erbium ions.

The experimental device utilized in this study is comprised of a z-cut LN crystal doped with 50~ppm of Erbium ions and a photonic integrated chip (PIC) containing two nano-fabricated silicon PhC cavities and a common coupler waveguide, nearly critically coupled to both PhC cavities (see Figs.~\ref{fig:setup}a and b). The PIC design allows spectral addressing of two clusters of REIs, which benefits spectral multiplexing. We fabricate the PICs separately from a silicon-on-insulator wafer with a 250~nm device layer and transfer it to the surface of the LN crystal using a pick-and-place technique~\cite{guo2022b}. LN is chosen because of its inversionless site symmetry, which allows linear Stark tuning of the Er$^{3+}$ ions' transition frequency~\cite{kaplyanskii2002,hastings-simon2006}. The symmetry furthermore determines the transition dipoles to be oriented along the x-y plane of the crystal. Ions near the LN surface therefore couple evanescently to the transverse electric (TE) field of the cavity. The PhC cavities on the LN substrate have a typical quality factor of $Q=50,000$, and finite-element simulations reveal mode volumes of 0.09~$\upmu$m$^3$ and an evanescent field strength at the Si-LN interface around 45\% of its maximal value. These parameters lead to a Purcell factor of 1,000 for a two-level atom at the Si-LN interface. Considering the branching ratio of 0.22~\cite{mcauslan2009}, the maximal enhancement of the Y$_1$ to Z$_1$ emission rate is expected to be 220. The device under test (DUT) is furthermore equipped with two parallel copper electrodes positioned above and below the 2~mm thick sample, which allows the creation of an electric field along the crystal c-direction. The coupler waveguide end is inverse tapered and sticks out of the LN sample. Light is coupled from and to a lensed fiber with typical fiber coupling efficiencies of around 60\%. The whole device is cooled inside a dilution refrigerator to around 20~mK (see the Supplemental Materials for more details).

\begin{figure}[t]
	\centering
	\includegraphics[width=\columnwidth]{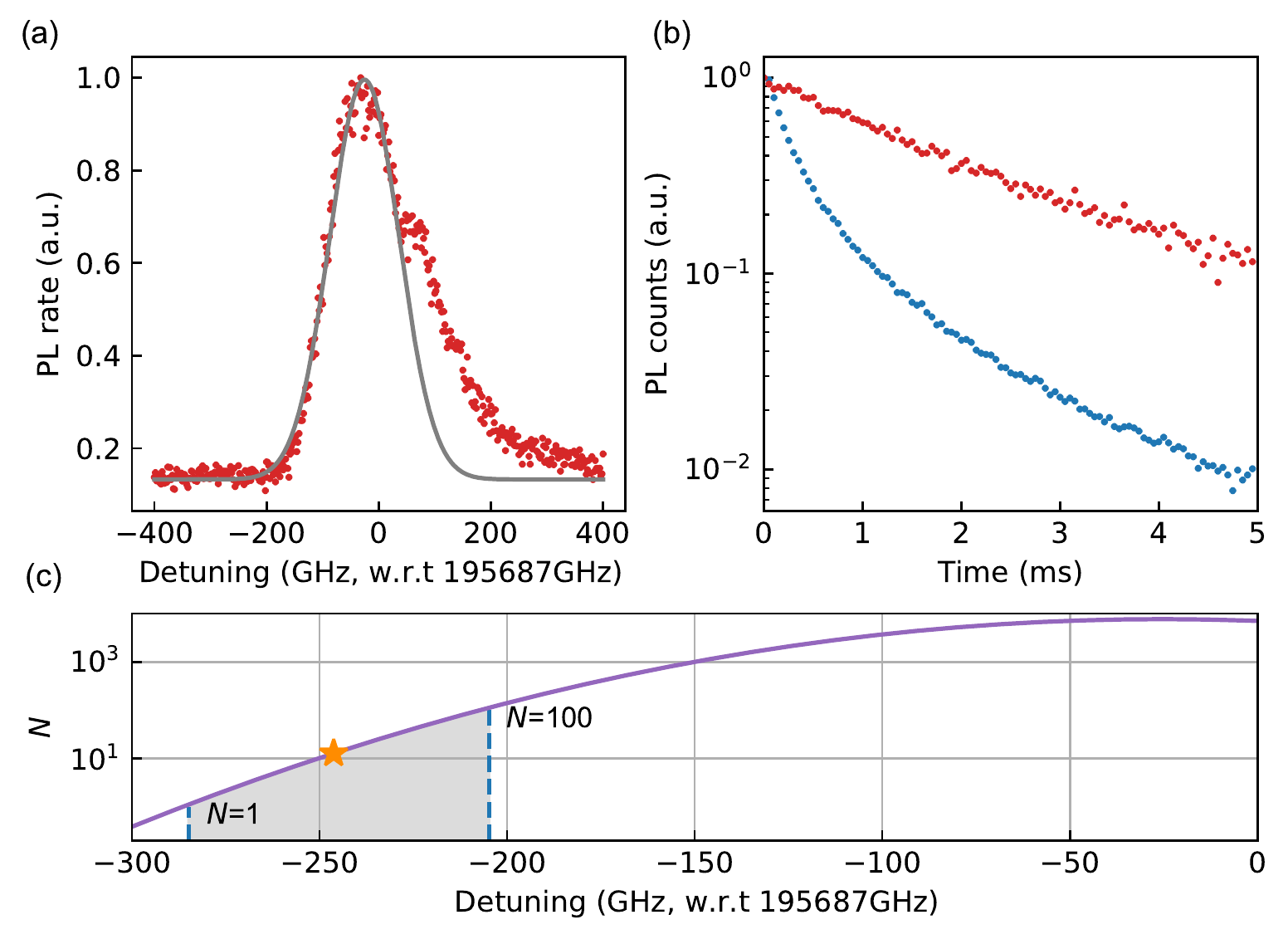}
	\caption{Characterization of ensemble Er$^{3+}$ ions. (a) PL spectrum of the Er ions under the coupler waveguide. The gray curve shows a Gaussian fit. (b) Time-resolved PL of the ions with the PhC cavity enhancement (blue, cavity resonance at 1532.5~nm) and without enhancement (red). (c) Simulated average cavity-addressable ion number $N$ within a frequency interval of 4~GHz as a function of the cavity resonance frequency. The star marks the frequency used in the single-ion measurements. The gray shaded area indicates the frequency region where spectrally resolved single ions are expected, with two blue dashed lines identifying $N=1$ and $100$.}
	\label{fig:spectrum}
\end{figure}

In a first step we characterize ensemble properties of Er:LN, starting with its inhomogeneously-broadened absorption line. Towards this end we employ photo luminescence (PL) excitation spectroscopy with the setup and the time sequence shown in Fig.~\ref{fig:setup}c and d, respectively. We excite ions near the Si-LN interface with a short optical pulse (typically 1~$\upmu$s) and record their emission using a superconducting nanowire single photon detector (SNSPD). Note that both the coupler waveguide and the PhC cavities couple to erbium ions. But while the coupler waveguide is very broadband and creates no Purcell enhancement, the PhC cavity only accepts a narrow spectral band and contributes to the Purcell enhancement. Note that the inhomogeneously broadened line of Er:LN is around 180~GHz~\cite{thiel2010}, greatly exceeding our cavity linewidth of around 4~GHz. Therefore, we choose a cavity with a resonance far away from the Er$^{3+}$ transition such that ions solely couple with the coupler waveguide. By scanning the excitation laser frequency (calibrated using a wavemeter with tested frequency precision of around 2~MHz) around the $Z_1-Y_1$ transition of Er$^{3+}$ at around 1532~nm, we obtain the spectrum shown in Fig.~\ref{fig:spectrum}a. The inhomogeneous line shows an asymmetric feature with a Gaussian profile and an additional bulge on the high-frequency side, arising from a phonon sideband transition~\cite{babajanyan2013a}. Using a Gaussian fit, we extract an inhomogeneous line centered at 1531.8~nm with 145.3~GHz width, in good agreement with the literature value. Combining the knowledge of the Er$^{3+}$ frequency distribution with the ion density and cavity mode volume, we calculate the average ion number $N$ within the PhC cavity linewidth as a function of the cavity resonance frequency (see Supplemental Materials). In order to spectrally resolve single ions within the cavity, one needs an average ion number per mode volume and resonance width between 1 and 100. The total frequency range for this is $\sim$80~GHz, as indicated in Fig.~\ref{fig:spectrum}c. 

Next, we pick a PhC cavity with resonance centered on the inhomogeneous line, which allows us to observe cavity-based enhancement of the ensemble decay rate, see Fig.~\ref{fig:spectrum}b. The PL signal shows a 250 times larger intensity and a significantly decreased decay time. The non-exponential decay profile is caused by contributions from different ions that experience different magnitudes of enhancement~\cite{zhong2017a}.

\begin{figure}[t]
	\centering
	\includegraphics[width=\columnwidth]{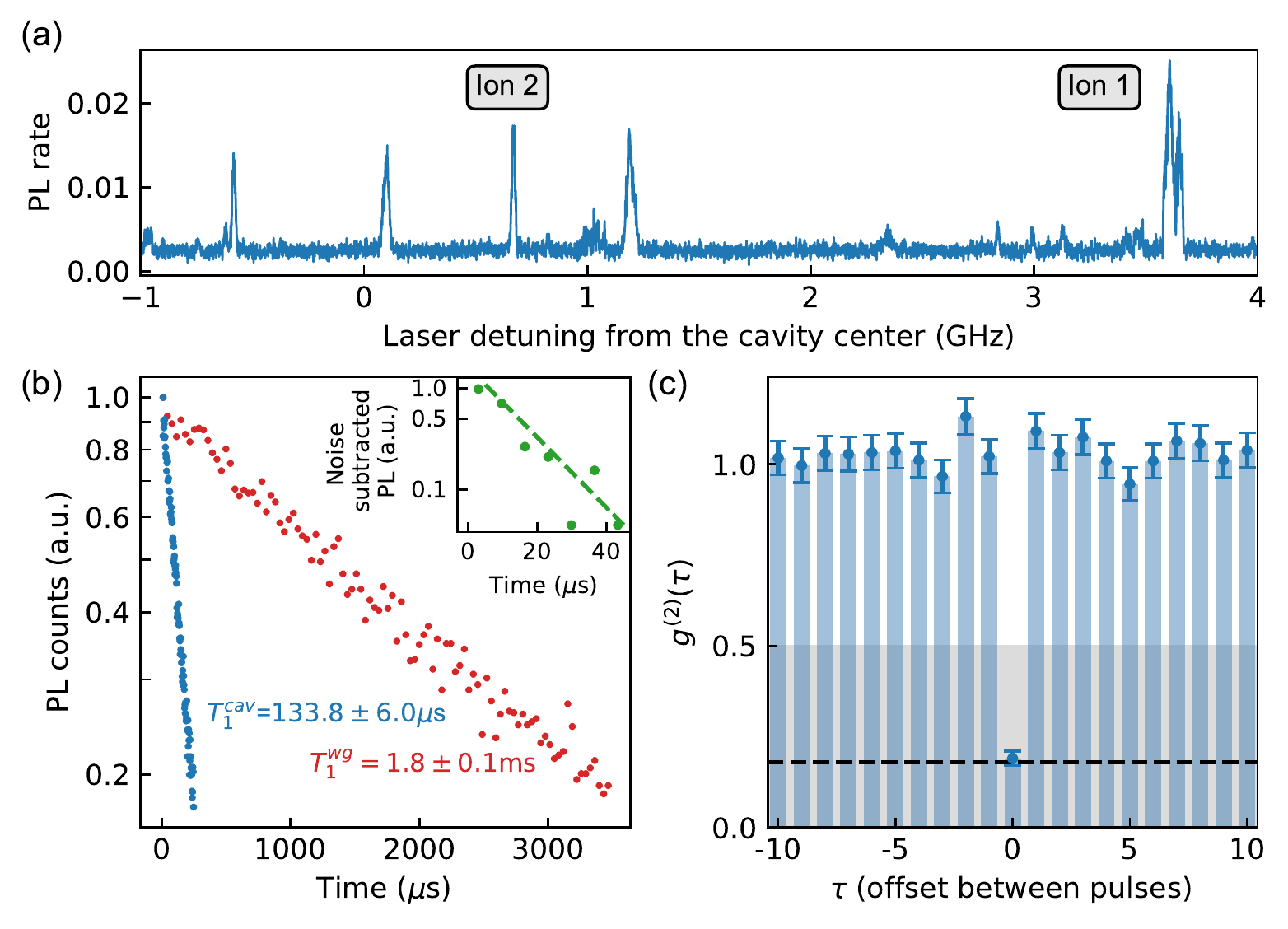}
	\caption{Characterization of single Er$^{3+}$ ions. (a) PL spectrum of ions coupled to a PhC cavity (resonance at 1533.8952~nm, linewidth of 3.8~GHz). The spectrum is measured with 1~MHz precision and each data point is accumulated for 6~s. (b) Time-resolved PL from cavity-enhanced Ion 1 (in blue). The PL decay of the Er$^{3+}$ ensemble that couples only to the waveguide (no Purcell enhancement) is shown for comparison (red). The inset shows an ion coupled to another PhC cavity with a decay constant of $12.5\pm 3.9~\upmu$s. (c) Second-order auto-correlation function $g^{(2)}(\tau)$ measured during 30~min with photons from Ion 1 using a Hanbury–Brown-Twiss setup. We group all photon detections during 90~$\upmu$s after each excitation pulse into a single time bin, and the horizontal axis shows the auto-correlation coefficient as a function of the difference between excitation pulses. The error bars represent one standard deviation. The grey shaded area indicates the single emitter regime and the black dashed line shows the dark count contribution.}
	\label{fig:hbt}
\end{figure}

As the second step, we look for spectrally isolated individual ions coupled to a PhC cavity. Towards this end, we choose a cavity with its resonance within the tail of the inhomogeneously broadened line to reduce the average number of ions within the cavity linewidth below 100. Specifically, we pick a cavity with resonance at 1533.8952~nm, two FWHM of the Er$^{3+}$ linewidth away from the inhomogeneous line center. The frequency is marked using a star in Fig.~\ref{fig:spectrum}c. Fig.~\ref{fig:hbt}a depicts five individual peaks within the cavity linewidth, which we interpret as spectrally isolated single ions. Some of the less prominent peaks in the spectrum may be a result of ions with reduced Purcell enhancement. Notably, the observed number of PL peaks agrees well with the estimated average ion number of 12 within the cavity resonance. We focus on the peak labeled Ion 1 in Fig.~\ref{fig:hbt}a for its highest brightness, and measure its time-resolved PL as shown in Fig.~\ref{fig:hbt}b. We extract an exponential decay lifetime of 133.8~$\upmu$s, 13-fold shorter than the waveguide measurement result. Measurements of additional PL peaks resulted in lifetimes as short as 12.5~$\upmu$s (see inset of Fig.~\ref{fig:hbt}b), corresponding to rate enhancements up to a factor of 143. These values are smaller than the expected value of 220, which we attribute to non-ideal positioning of the ions in the cavity field. Better-positioned ions could have been found by tuning the cavity frequency by means of gas tuning~\cite{dibos2018}. 

To confirm that the isolated PL peaks originate indeed from single Er$^{3+}$ ions, we now measure the second-order auto-correlation coefficient $g^{(2)}(0)$ of Ion 1. Its value of 0.19$\pm$0.02 extracted from Fig.~\ref{fig:hbt}c is significantly lower than the threshold of $0.5$ for two emitters~\cite{gerry2004}, indicating that the majority of the detected photons stem from a single ion. We obtain similar results for other ions. For more insights we analyze the detector noise contributions to the coincidence count rates, finding that accidental coincidences between detected photons and dark counts fully account for the imperfect $g^{(2)}(0)$ (see the Supplemental Materials). To further improve the single-photon purity towards $g^{(2)}(0)=0$, a higher signal-to-noise ratio is therefore required. This can be achieved by increasing the Purcell factor, which will improve the photon emission rate, and by optimizing the system efficiency.

\begin{figure}[t]
	\centering
	\includegraphics[width=\columnwidth]{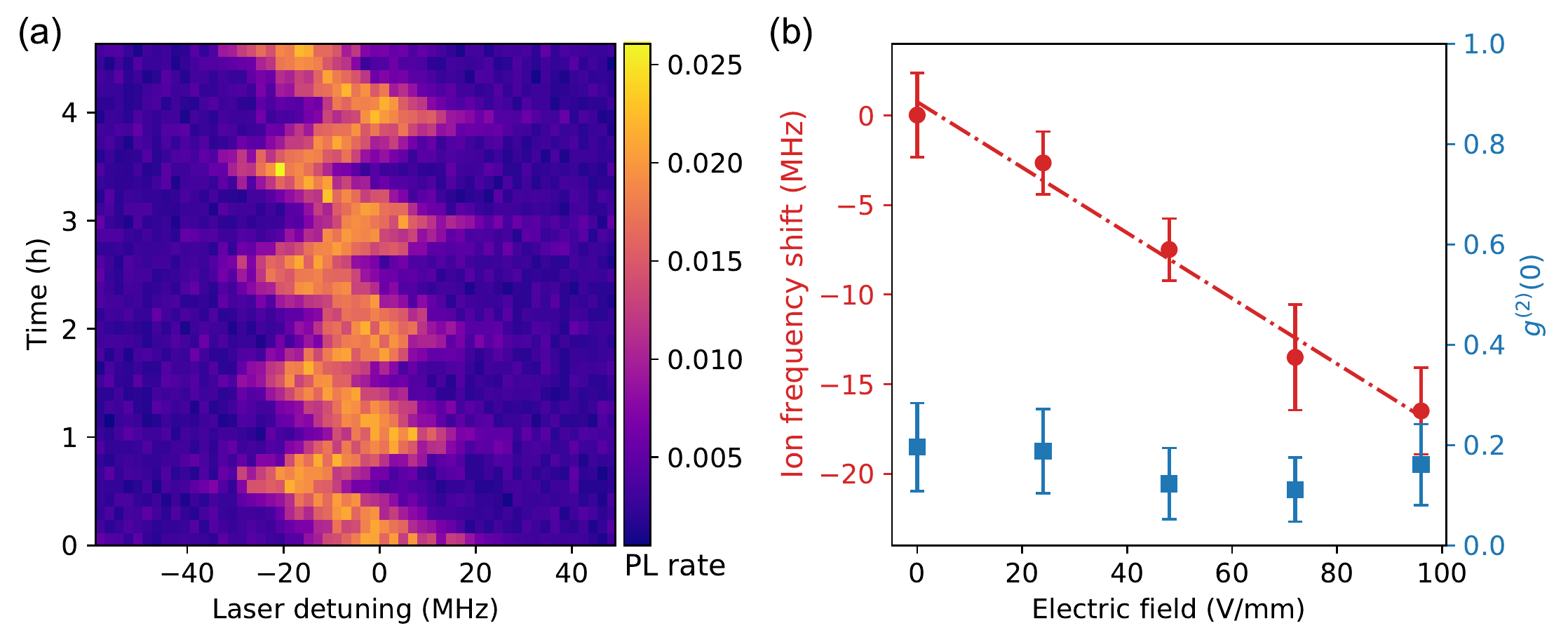}
	\caption{Linear Stark tuning of a single Er$^{3+}$ ion. (a) The spectral line of a single Er$^{3+}$ ion shifts back and forth when a triangular voltage ramp is applied across the Er:LN crystal. (b) Frequency shift and auto-correlation coefficient $g^{(2)}(0)$ of emitted photons as a function of the electric field amplitude. The red dashed line is a linear fit to the frequency shift of the ion; error bars are calculated from data taken during different ramps. The error of the correlation coefficient is calculated from Poissonian photon counting statistics.}
	\label{fig:stark}
\end{figure}

We now tune the frequency of the Er$^{3+}$ ions by applying an external DC electric field to the electrodes. This shifts the energy levels of emitters with a permanent dipole moment, which is known as the linear DC Stark effect~\cite{kaplyanskii2002}. When the dipole moments of the ground and of the excited states experience different shifts, the external electric field leads to a shift of the associated optical transition frequency. We focus on Ion 2 (for its reduced spectral width compared to Ion 1) and observe its frequency shift while cycling the voltage from 0~V to 640~V multiple times in a linear fashion. The results are depicted in Fig.~\ref{fig:stark}. We extract a Stark coefficient of $182.9$~kHz/V~mm$^{-1}$, not far from the literature value of 250~kHz/V~mm$^{-1}$~\cite{hastings-simon2006}. We attribute the small discrepancy to imperfect modelling of the electric field.

To confirm that the single-ion nature is not affected by the electric field, we furthermore measure the auto-correlation function under different electric field amplitudes. As shown in Fig.~\ref{fig:stark}b, $g^{(2)}(0)$ remains constant, close to the noise-limited threshold. This establishes Stark tuning as a useful tool for controlling the frequency of single rare-earth ions in quantum applications.

\begin{figure}[t!]
	\centering
	\includegraphics[width=1\columnwidth]{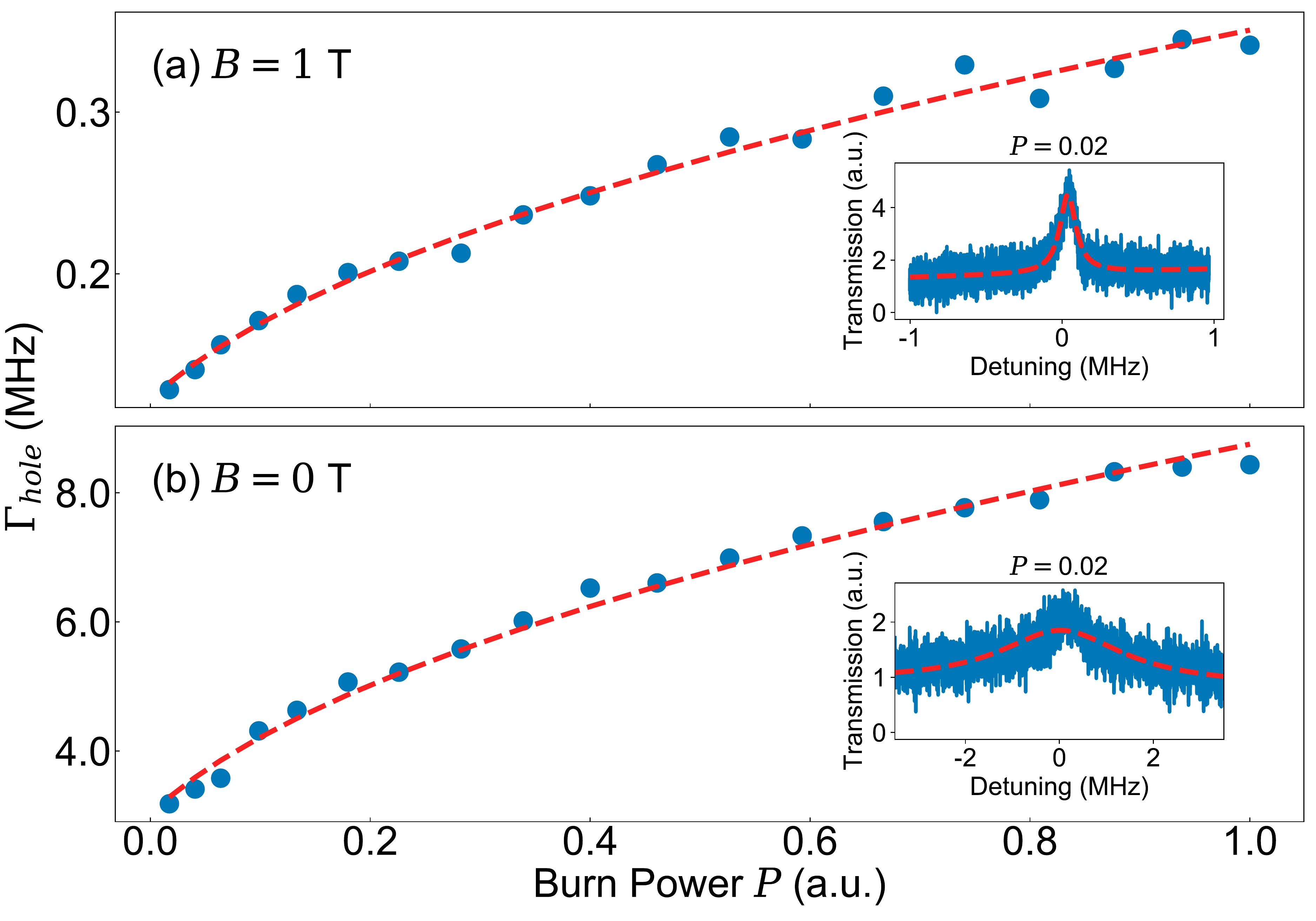}
	\caption{ (a) A spectral hole of 130~kHz width ``burned" with minimum power into the inhomogeneously-broadened absorption line of Er:LN at $T=1.1$~K and $B=1$~T (inset), and hole widths $\Gamma_\textrm{hole}$ as a function of burning power (main graph). For comparison, panel (b) shows a 3.2~MHz-wide hole created under the same conditions but at $B=0$~T. Extrapolating the data in the main graphs to zero burning power, we find power broadening-free hole widths of $122\pm4$~kHz and $3.02\pm0.1$~MHz. Error bars are smaller than the size of each data point.}
	\label{fig:SHB}
\end{figure}

Finally, we give a brief outlook on the possibility of creating emitters in different---potentially distant---crystals that emit indistinguishable photons. This implies (i) lifetime (Fourier)-limited, and (ii) identical spectra. 
Fig.~\ref{fig:hbt}a shows that the measured line widths of the observed PL peaks range between 10 and 35~MHz, largely exceeding, e.g., the Fourier limit of around 25~kHz imposed by the (reduced) lifetime of 12.5~$\upmu$s (we used $\Gamma_{hole}=1/\pi T_1$). We attribute this excess broadening to laser frequency fluctuations as well as ion spectral diffusion caused by spin flip-flops~\cite{thiel2010}. Spectral diffusion can be largely removed by polarizing the spin reservoir using a strong magnetic field~\cite{thiel2010}. Indeed, at a temperature of 1.1~K, a magnetic field of 1~T, and after a delay of 380~$\upmu$s, we find that the width of a spectral hole, which is directly related to the width of a PL peak, burned into the inhomogeneously broadened erbium absorption line decreases from roughly 3~MHz to 130~kHz (see Fig.~\ref{fig:SHB}). Significant additional narrowing is expected at a higher magnetic field~\cite{thiel2010} and at a temperature of a few tens of mK, where coherence times, measured using two-pulse photon echoes, up to 180~$\upmu$s have been reported~\cite{wang2022j}. We therefore conjecture that, under readily achievable experimental conditions, the linewidth of an isolated PL peak becomes indeed Fourier limited. We also note that the lifetime can be further decreased (and hence the related spectral width increased) by choosing an ion at an optimized location with a maximum electric field, by decreasing the mode-volume of the cavity, or by increasing its $Q$ factor. We are therefore confident that it is possible to create lifetime-limited photons using individual erbium ions. Additional spectral diffusion can be compensated using Stark-shift-based feedback, similar to what has been demonstrated with quantum dots using strain or electric fields~\cite{patel2010,flagg2010}.

Such Stark tuning also allows matching the spectra of two ions that are located in different crystals, thereby creating indistinguishable photons. This enables multi-photon interference, which is key to quantum network operations based on Bell-state measurements, such as quantum teleportation. Please note that due to the simple design of the electrodes in our proof-of-principle demonstration, the observed Stark shift is only slightly larger than the width of the ion's spectral line. However, much larger Stark tuning, up to Gigahertz using only 20~V, can be expected by introducing nanoscale on-chip electrodes, for which the fabrication process is well-developed and directly compatible with our fabrication process~\cite{yang2023}.

In summary, we have detected single photons from individual erbium ions using Purcell enhancement through a photonic crystal cavity. Additionally, we have demonstrated linear Stark tuning of the ion frequency without altering its single-photon emission statistics. By exploiting the piezo-optic effect of lithium niobate, it is furthermore possible to tune the cavity resonance, making Er:LN a versatile quantum emitter platform. However, to achieve lifetime-limited linewidth, a quieter spin bath and a more stable charge environment are required. The former can be created through low-dose ion implantation and by polarizing the spin bath using a large magnetic field, while the latter can be achieved by implementing Stark tuning-based feedback. This will allow dynamical compensation of spectral diffusion as well as spectral tunability, and thereby open the door to rare-earth-ion-based quantum networks.

\textit{Acknowledgments}. We would like to thank Robert Stockill, Bas Hensen and Antariksha Das for valuable discussions, as well as Raymond Schouten and Roy Birnholtz for their help during various stages of the experiments. We further acknowledge assistance from the Kavli Nanolab Delft. This work is financially supported by the European Research Council (ERC CoG Q-ECHOS, 101001005), and by the Netherlands Organization for Scientific Research (NWO/OCW), as part of the Frontiers of Nanoscience program, as well as through Vrij Programma (680-92-18-04) and Klein (OCENW.KLEIN.555) grants. Y.Y. gratefully acknowledges funding from the European Union under a Marie Sk\l{}odowska-Curie fellowship. No parts of this paper were written using ChatGPT.

\clearpage

\renewcommand{\andname}{\ignorespaces}

\setcounter{figure}{0}
\setcounter{table}{0}
\setcounter{equation}{0}

\onecolumngrid

\global\long\def\theequation{S\arabic{equation}}
\global\long\def\thefigure{S\arabic{figure}}
\renewcommand{\thetable}{S\arabic{table}}
\renewcommand{\arraystretch}{0.6}
\renewcommand{\theHfigure}{S\arabic{figure}}

\normalsize

\begin{center}
	\linespread{1.5}
	\Large{\textbf{Supplemental Material}}\\ \large{\textbf{Frequency tunable, cavity-enhanced single erbium quantum \\ emitter in the telecom band}}
\end{center}

\section{Photonic crystal cavity simulation}
\begin{figure}[b]
	\centering
	\includegraphics[width=0.8\textwidth]{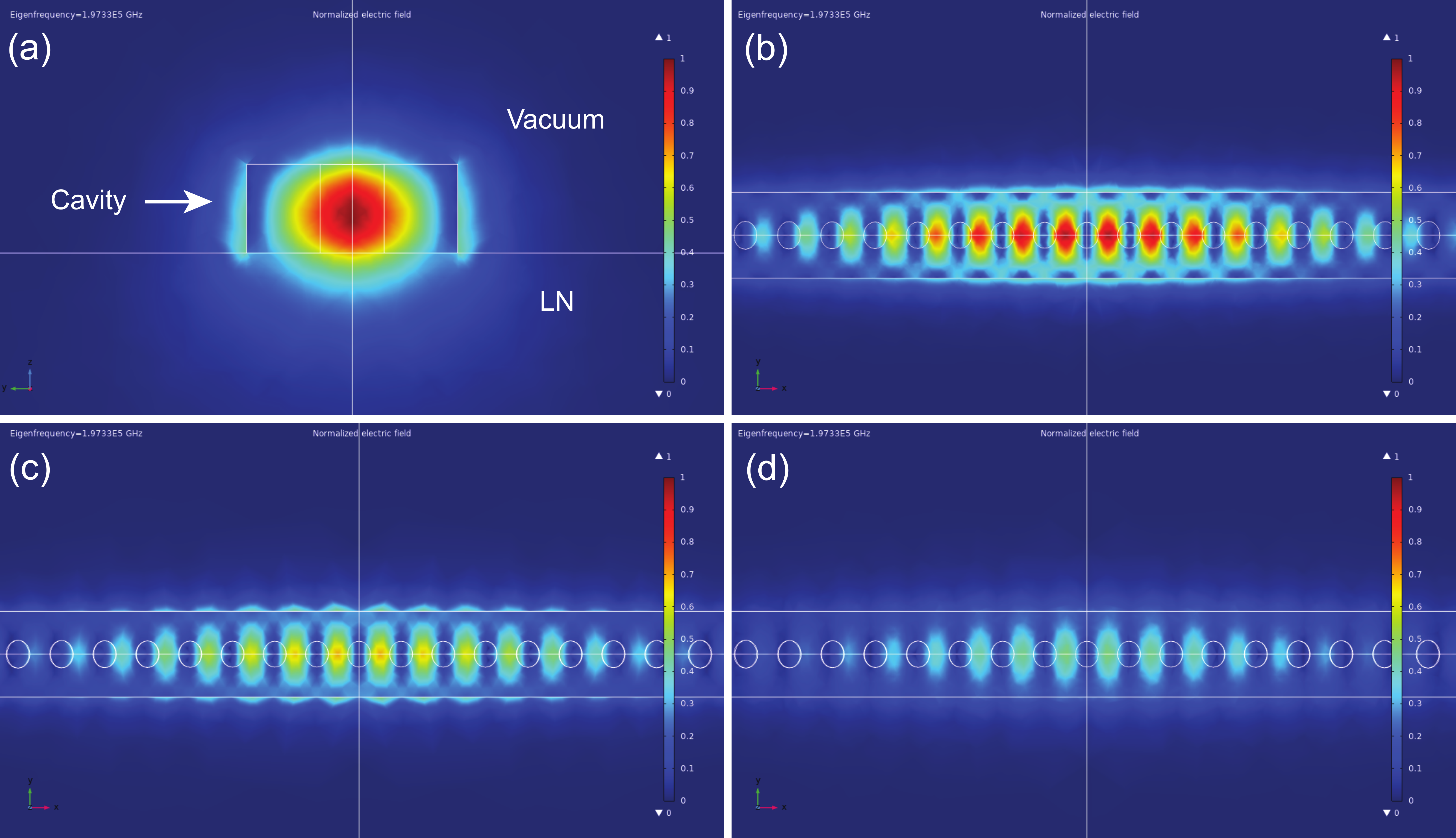}
	 \caption{Simulated electric field distribution of the PhC cavity. (a) Radial cross-section near the cavity center. (b), (c), and (d) show the result in the planes 125~nm above the Si-LN interface (the center of the silicon), at the Si-LN interface, and 50~nm below the interface, respectively.}
	\label{fig:simulation}
\end{figure}

We perform finite element simulation of a one-dimensional silicon (refractive index $n_{\text{Si}}=3.45$) photonic crystal cavity on top of a lithium niobate crystal (LN, refractive index $n_{\textrm{LN}}=2.30$) using  COMSOL. Holes at the end of the silicon beam create a bandgap for optical light with a wavelength of around 1532~nm. The variation of holes towards the center of the beam perturbs the bandgap, creating optical resonances. The simulation results in a fundamental optical mode with the electric field distribution shown in Fig.~\ref{fig:simulation}. We get a nominal mode volume of 
\begin{equation}
    V_{\textrm{mode}} = \frac{\int\epsilon |E(\textbf{r})|^2\textrm{d}V}{\epsilon| E_{\textrm{max}}|^2} = 0.09~\upmu \textrm{m}^3,
\end{equation}
where $\epsilon$ is the permittivity of the corresponding medium, and a quality factor of $Q_{\textrm{sim}}=6\times 10^5$. We attribute the decrease of the measured quality factor $Q_{\textrm{meas}}\approx 5\times 10^4$ to imperfection in nanofabrication. The optical mode leaks around 10\% into the lithium niobate substrate and the electric field at the interface is 45\% of its maxima at the center of the silicon cavity.

We estimate the Purcell factor of our device using
\begin{equation}
    F_p = \frac{3}{4\pi^2}\left(\frac{\lambda}{n_{\textrm{LN}}}\right)^3 \cdot \frac{Q}{V_{\textrm{mode}}}\cdot\Bigl|\frac{E(\textbf{r})}{E_{\text{max}}}\Bigr|^2\cdot\frac{\beta}{\chi_L},
\end{equation}
where $\beta=0.22$ is the branching ratio of the corresponding $Z_1-Y_1$ transition and $\chi_L = [(n_{LN}^2+2)/3]^2$ is the local field correction. Substituting the electric field at the Si-LN interface, we predict $F_p = 220$.

\section{Device fabrication and assembly}
\begin{figure}[h]
	\centering
	\includegraphics[width=\textwidth]{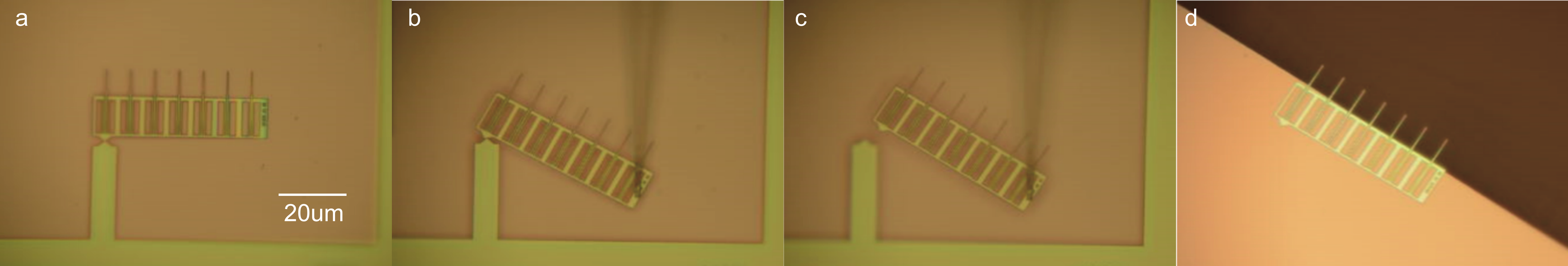}
	 \caption{Pick-and-place process. (a) A group of silicon PhC cavity devices, which is only weakly connected to the surroundings by a thin tether after the buried oxide layer is removed. (b) A tapered fiber attaches to the group of devices and breaks the thin tether. (c) The group adheres to the fiber, and is brought to the target. (d) The cavities are placed on the surface of a lithium niobate crystal and aligned to the edge.}
	\label{fig:slapping}
\end{figure}

The PhC cavities are fabricated from a silicon-on-insulator (SOI) wafer with a 250~nm device layer and a 3~$\upmu$m buried oxide layer. In order to transfer the devices to the lithium niobate surface, we group seven nanobeams together as shown in Fig.~\ref{fig:slapping}a, and isolate the device from the majority of its surroundings with only a thin tether for support. We pattern the device structure using electron beam lithography followed by HBr/Ar reactive ion etching. We perform a piranha cleaning step and finally remove the buried oxide layer beneath the device layer with a vapor hydrofluoric acid (HF) either in liquid or vapor form. For the liquid wet etching case, we use critical point drying to protect the fragile tethers during the final drying step. The average yield of the fabrication process is about 70\%.

We use a pick-and-place technique~\cite{guo2022b_sm} to transfer the silicon PhC cavities onto the surface of the lithium niobate. The workflow is shown in Fig.~\ref{fig:slapping}. We start by sticking a tapered fiber with the help of van der Waals and electrostatic forces to the group of PhC cavities. We then break the tether by moving the sample stage. The device attached to the tapered fiber tip is brought above the lithium niobate, aligned with the edge and placed on the surface. The device's adherence to the lithium niobate is stronger than to the fiber due to a much larger contact surface, enabling the detachment of the fiber. All movements are performed by motorized stages under an optical microscope.

\section{Average accessible ions per cavity}

The average number of accessible ions per cavity resonance width can be expressed as,
\begin{equation}
    N = V_{\textrm{cav}}\cdot \rho \cdot D(f).
\end{equation}
$V_{\textrm{cav}}$ represents the size of the `cavity regime' within the LN, where ions experience a significant enhancement. We estimate it by integrating the space in the LN where the cavity field instensity is greater than $1/e^2$ of its maximum and get $V_{\textrm{cav}}\approx 4.5\times 10^{-19}$~m$^3$. $\rho$ is the ion density in the LN crystal and $D(f)$ is the ion frequency distribution function (given in Fig.~\ref{fig:spectrum}a in the main text). Considering the 50~ppm nominal concentration of ions and the inhomogeneously broadened line shape, we calculate $N$ as a function of the cavity detuning, see Fig.~\ref{fig:spectrum}c in the main text. We find $N=12.7$ for the specific cavity used in the single ion measurements.

\section{Noise analysis of the auto-correlation measurement}

We consider the contribution of noise caused during photon deteection  to the auto-correlation measurement. It includes dark counts of the superconducting nanowire single-photon detector (SNSPD) and background light, and can be modeled as a classical light source with auto-correlation coefficient of $\braket{I_n^2}/\braket{I_n}^2=1$. The noise is mixed with the photo luminescence (PL) from a single ion, which is assumed to feature ideal antibunching with $\braket{I_{ion}^2}/\braket{I_{ion}}^2=0$. We can estimate the measured auto-correlation coefficient $g^{(2)}(0)$ using
\begin{equation}\label{eq:g2noise}
    g^{(2)}(0)=\frac{ \braket{I^2}}{ \braket{I}^2}=\frac{\braket{\left(I_{ion}+I_{n}\right)^2}}{\braket{ I_{\text {ion }}+I_{n}}^2}=\frac{\braket{ I_{i o n}^2}+\braket{ I_{n}^2}+2\braket{ I_{ion }}\braket{ I_{n}}}{\braket{ I_{i o n}}^2+\braket{ I_{n}}^2+2\braket{ I_{i o n}}\braket{ I_{n}}}=\frac{2\text{SNR}+1}{(\text{SNR}+1)^2},
\end{equation}
where the signal-to-noise ratio SNR is defined as $\text{SNR}=\braket{I_{ion}}/{I_n}$. 

In our experiment, the measured detector noise in a 90~$\upmu$s window is $1.15\times10^{-3}$ (corresponding to 12.7~Hz) and the signal intensity is $2.20\times10^{-2}$, leading to $\text{SNR}=19.2$. Substituting the $\text{SNR}$ into Eq.~\ref{eq:g2noise}, we find $g^{(2)}(0)=0.1797$, in good agreement with our measurement result.

\section{Ensemble spectroscopy}

To shed some light on the width of the single PL emission lines, which is around 10~MHz over around 1~min (see Fig.~\ref{fig:hbt}a in the main text), we perform standard spectral hole burning (SHB) and 2-pulse photon echo (2PPE) experiments using an identical crystal. The experiments are conducted in another cryostat that allows the application of a magnetic field along the crystal c-axis, but whose base-temperature is limited to 1.1K -- much higher than the 20 mK in the dilution refrigerator that is used for all other measurements reported in this paper. As such, the results  provide only indicative bounds for the relevant properties, with an in-depth investigation being beyond the scope of this paper and remaining for future experiments.

\textbf{Spectral hole burning.} For SHB experiments~\cite{kaplianskii1987,macfarlane2002}, resonant, narrow-band laser light interacts during some time with a spectrally narrow  subensemble of  ions and pumps them to the excited state. This ``burn" process leads to a modification of the transmission at the frequencies at which the ions can absorb -- either during a time limited by the excited state lifetime of the ions, around 1.8~ms in the case of erbium in LiNbO$_3$, or, in case a magnetic field is applied, a time limited by the relaxation time between different spin levels (so-called persistent spectral holes), e.g. several minutes at $B=1.9$~T and $T\approx0.7$~K in the case of Er:LiNbO$_3$~\cite{askarani2019}. 

As shown in \ref{fig:SHB}.~5 in the main text, we vary the excitation power for two different magnetic fields (0~T, 1~T) over two orders of magnitude, with the smallest value chosen such that we can barely see a spectral hole. This allows extrapolating the measured hole width to zero excitation power where it is not limited by power broadening:
\begin{equation}
    \Gamma_{hole}=\Gamma_{hom}(1+\sqrt{1+P/P_{Sat}}).
\label{eq:SHB}
\end{equation}
\noindent
Here $P_{Sat}$ is the saturation power and $\Gamma_{hom}$ is the homogeneous linewidth \cite{lethokov1976}. Eq.~\ref{eq:SHB} assumes small optical depth, a correct assumption in our case. The ``burn" time is set to 50~$\mu$s, and the time until the moment when the subsequent ``read" pulse, which is chirped during 500 $\mu$s over 10 MHz, scans over the spectral hole is 380~$\mu$s. All measurements are done at a wavelength of 1532~nm, i.e.\ at the center of the inhomogeneously broadened absorption line of Er:LiNbO$_3$.

\begin{figure*}[b!]
	\centering
	\includegraphics[width=0.6\columnwidth]{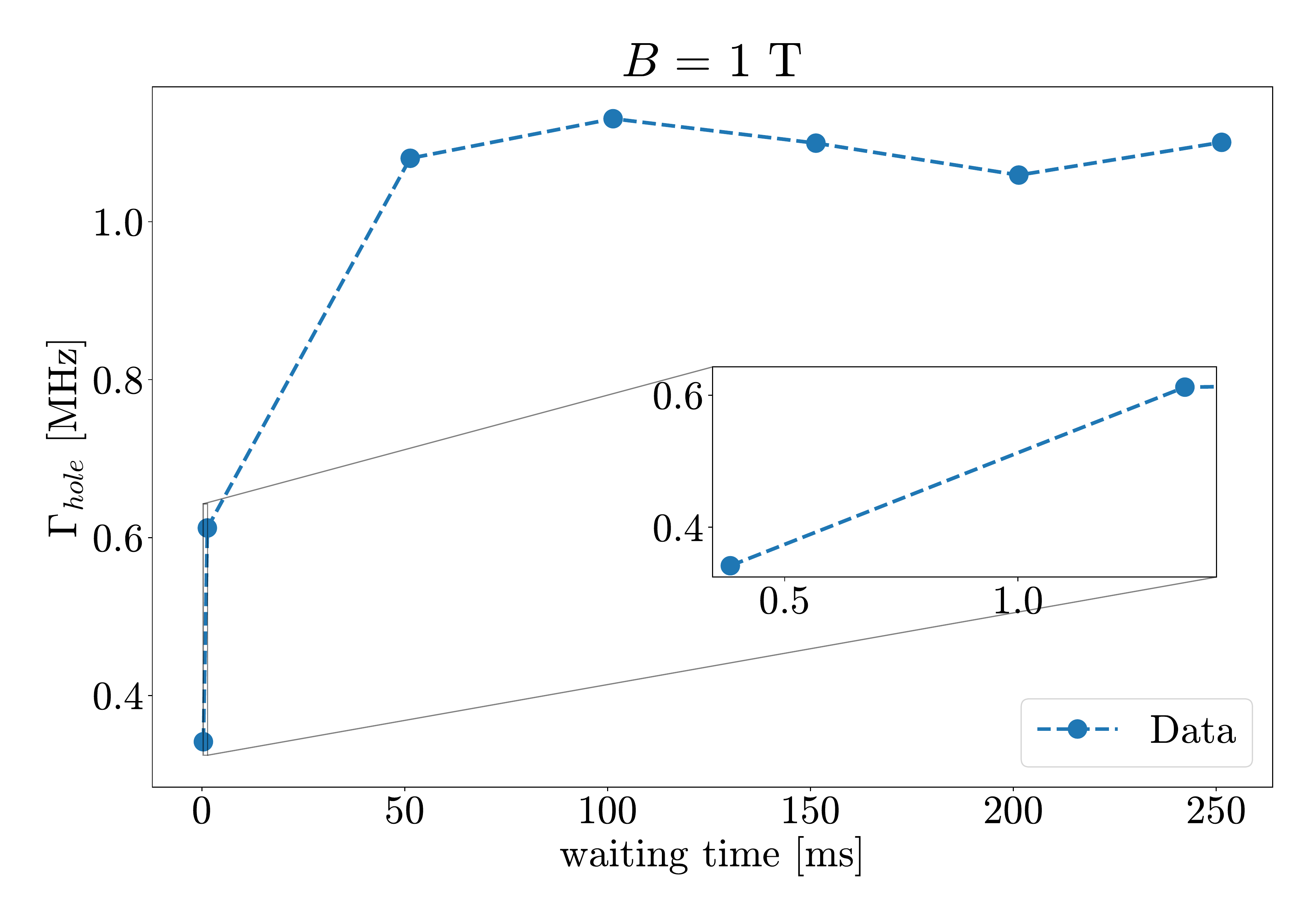}
	 \caption{Hole broadening for $B=1$~T as a function of waiting time. The holes were created with the largest burning power used in the measurements depicted in Fig.~\ref{fig:SHB} in the main text. Error bars are smaller than the size of each experimental data point.}
  \label{fig:SHBSM}
  \end{figure*}

Fig.~5 in the main text shows the evolution of the width of the spectral hole for the two magnetic fields as the excitation power is varied. We would like to note several things:

\begin{itemize}
\item For $B=0$~T, the minimum holewidth is around 3~MHz. This is not far from the width of the PL peak in Fig.~3a of the main text. The difference is due to spectral diffusion caused by erbium spin-flips \cite{thiel2010_sm}.

\item For $B=1$~T the hole width narrows significantly, as was also observed in~\cite{thiel2010_sm}. 

\item{As shown in Fig.~\ref{fig:SHBSM}, the hole width at $B=1$~T broadens slightly over time, reaching its maximum value of approximately 1~MHz after 50~ms.}

\item Further line narrowing is expected for higher magnetic fields of around 5~T, where spectral diffusion will be dominated only by nuclear spin flips \cite{thiel2010_sm}. Another possibility is to decrease the temperature below 1~K --  similar to results of coherence measurements using 2-pulse photon echoes, where the coherence (or memory) time, measured at $B=0.5$~T, increased from 90~$\mu$s at 1.1~K (see paragraph about 2-pulse photon echoes) to 180~$\mu$s at 20~mK~\cite{Wang2022}.

\end{itemize}

\begin{figure*}[htbp!]
	\centering
	\includegraphics[width=0.6\columnwidth]{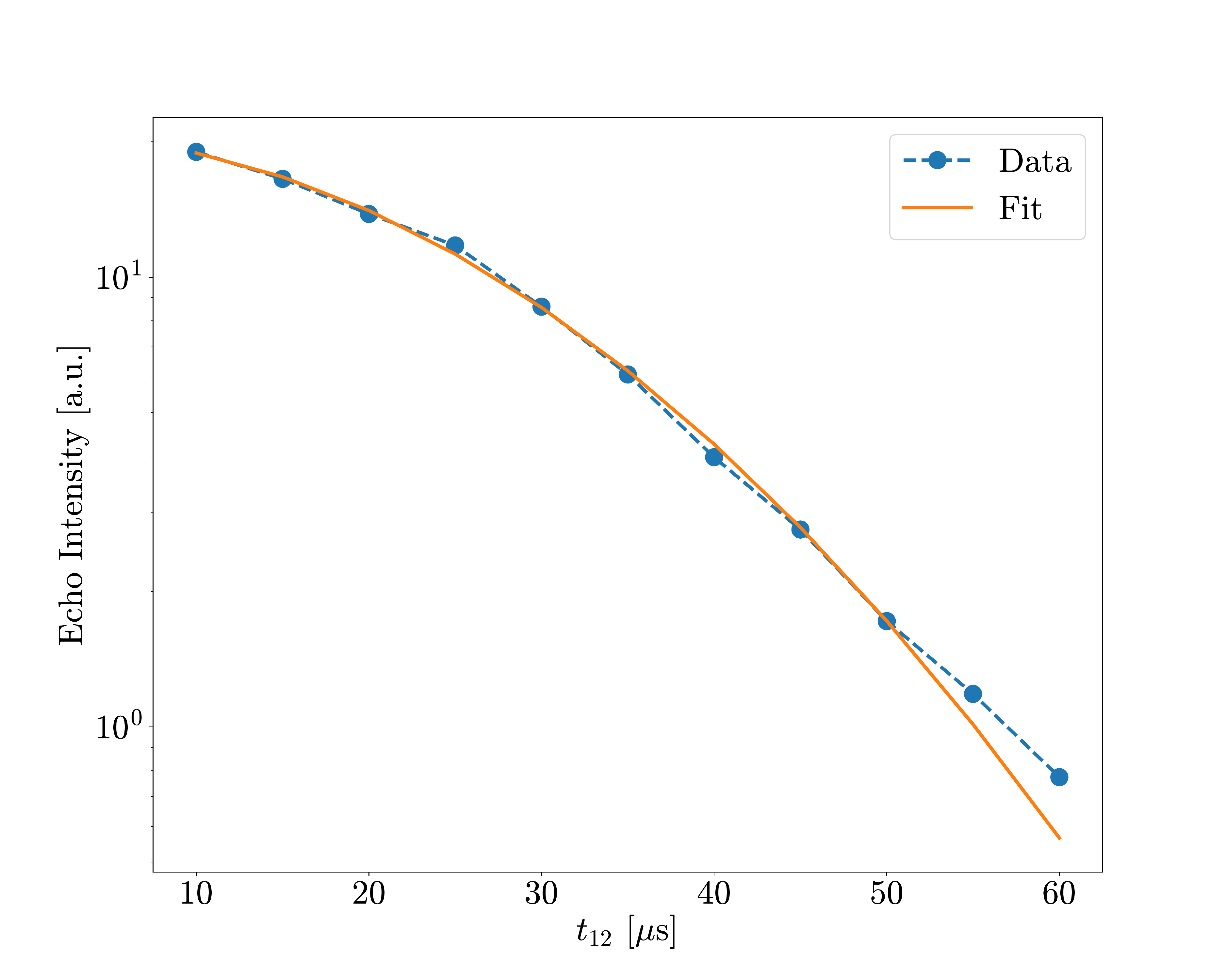}
	 \caption{2-pulse photon echoes. The variation of the waiting time between the first and the second pulse creates an echo with a super-exponentially decaying amplitude. A best fit results in a memory time $T_M$ of $89.7\pm 0.92$~$\mu$s and a spectral diffusion (Mims) coefficient $m=2.023\pm 0.083$. Error bars are smaller than the size of each experimental data point.}
  \label{fig:2PPE}
\end{figure*}

\textbf{2-pulse photon echoes.} 
In 2PPE experiments \cite{kaplianskii1987,macfarlane2002}, two resonant excitation pulses---ideally with pulse areas of $\pi/2$ and $\pi$---are sent with varying delay $t_{12}$ into the crystal, leading to the emission of an echo time $t_{12}$ after the second pulse. The decay of the echo intensity as a function of $t_{12}$ yields information about coherence and spectral diffusion measured on a shorter time-scale than SHB.

Fig.~\ref{fig:2PPE} depicts our results for a magnetic field of 1~T. As above, the laser was tuned to 1532~nm wavelength. To fit the data, we use the standard Mims expression~\cite{thiel2010_sm}

\begin{equation}
    I=I_0 e^{-2(2t_{12}/T_M)^m}
\end{equation}
\noindent
where T$_M$ is the memory time (effective coherence time, which relates to the effective homogeneous linewidth as $\Gamma_{hom}\sim 1/\pi T_M$) and $m$ is the Mims coefficient ($m$ is 1 if no spectral diffusion takes place and larger otherwise). We find a memory time of 89.7~$\mu$s.


\begin{thebibliography}{51}%
	\makeatletter
	\providecommand \@ifxundefined [1]{%
	 \@ifx{#1\undefined}
	}%
	\providecommand \@ifnum [1]{%
	 \ifnum #1\expandafter \@firstoftwo
	 \else \expandafter \@secondoftwo
	 \fi
	}%
	\providecommand \@ifx [1]{%
	 \ifx #1\expandafter \@firstoftwo
	 \else \expandafter \@secondoftwo
	 \fi
	}%
	\providecommand \natexlab [1]{#1}%
	\providecommand \enquote  [1]{``#1''}%
	\providecommand \bibnamefont  [1]{#1}%
	\providecommand \bibfnamefont [1]{#1}%
	\providecommand \citenamefont [1]{#1}%
	\providecommand \href@noop [0]{\@secondoftwo}%
	\providecommand \href [0]{\begingroup \@sanitize@url \@href}%
	\providecommand \@href[1]{\@@startlink{#1}\@@href}%
	\providecommand \@@href[1]{\endgroup#1\@@endlink}%
	\providecommand \@sanitize@url [0]{\catcode `\\12\catcode `\$12\catcode
	  `\&12\catcode `\#12\catcode `\^12\catcode `\_12\catcode `\%12\relax}%
	\providecommand \@@startlink[1]{}%
	\providecommand \@@endlink[0]{}%
	\providecommand \url  [0]{\begingroup\@sanitize@url \@url }%
	\providecommand \@url [1]{\endgroup\@href {#1}{\urlprefix }}%
	\providecommand \urlprefix  [0]{URL }%
	\providecommand \Eprint [0]{\href }%
	\providecommand \doibase [0]{https://doi.org/}%
	\providecommand \selectlanguage [0]{\@gobble}%
	\providecommand \bibinfo  [0]{\@secondoftwo}%
	\providecommand \bibfield  [0]{\@secondoftwo}%
	\providecommand \translation [1]{[#1]}%
	\providecommand \BibitemOpen [0]{}%
	\providecommand \bibitemStop [0]{}%
	\providecommand \bibitemNoStop [0]{.\EOS\space}%
	\providecommand \EOS [0]{\spacefactor3000\relax}%
	\providecommand \BibitemShut  [1]{\csname bibitem#1\endcsname}%
	\let\auto@bib@innerbib\@empty
	\bibitem [{\citenamefont {Kimble}(2008)}]{kimble2008}%
	  \BibitemOpen
	  \bibfield  {author} {\bibinfo {author} {\bibfnamefont {H.~J.}\ \bibnamefont
	  {Kimble}},\ }\bibfield  {title} {\bibinfo {title} {The quantum internet},\
	  }\href {https://doi.org/10.1038/nature07127} {\bibfield  {journal} {\bibinfo
	  {journal} {Nature}\ }\textbf {\bibinfo {volume} {453}},\ \bibinfo {pages}
	  {1023} (\bibinfo {year} {2008})}\BibitemShut {NoStop}%
	\bibitem [{\citenamefont {Wehner}\ \emph {et~al.}(2018)\citenamefont {Wehner},
	  \citenamefont {Elkouss},\ and\ \citenamefont {Hanson}}]{wehner2018}%
	  \BibitemOpen
	  \bibfield  {author} {\bibinfo {author} {\bibfnamefont {S.}~\bibnamefont
	  {Wehner}}, \bibinfo {author} {\bibfnamefont {D.}~\bibnamefont {Elkouss}},\
	  and\ \bibinfo {author} {\bibfnamefont {R.}~\bibnamefont {Hanson}},\
	  }\bibfield  {title} {\bibinfo {title} {Quantum internet: {{A}} vision for the
	  road ahead},\ }\href {https://doi.org/10.1126/science.aam9288} {\bibfield
	  {journal} {\bibinfo  {journal} {Science}\ }\textbf {\bibinfo {volume}
	  {362}},\ \bibinfo {pages} {eaam9288} (\bibinfo {year} {2018})}\BibitemShut
	  {NoStop}%
	\bibitem [{\citenamefont {Sangouard}\ \emph {et~al.}(2011)\citenamefont
	  {Sangouard}, \citenamefont {Simon}, \citenamefont {{de Riedmatten}},\ and\
	  \citenamefont {Gisin}}]{sangouard2011}%
	  \BibitemOpen
	  \bibfield  {author} {\bibinfo {author} {\bibfnamefont {N.}~\bibnamefont
	  {Sangouard}}, \bibinfo {author} {\bibfnamefont {C.}~\bibnamefont {Simon}},
	  \bibinfo {author} {\bibfnamefont {H.}~\bibnamefont {{de Riedmatten}}},\ and\
	  \bibinfo {author} {\bibfnamefont {N.}~\bibnamefont {Gisin}},\ }\bibfield
	  {title} {\bibinfo {title} {Quantum repeaters based on atomic ensembles and
	  linear optics},\ }\href {https://doi.org/10.1103/RevModPhys.83.33} {\bibfield
	   {journal} {\bibinfo  {journal} {Rev.\ Mod.\ Phys.}\ }\textbf {\bibinfo
	  {volume} {83}},\ \bibinfo {pages} {33} (\bibinfo {year} {2011})}\BibitemShut
	  {NoStop}%
	\bibitem [{\citenamefont {Reiserer}\ and\ \citenamefont
	  {Rempe}(2015)}]{reiserer2015}%
	  \BibitemOpen
	  \bibfield  {author} {\bibinfo {author} {\bibfnamefont {A.}~\bibnamefont
	  {Reiserer}}\ and\ \bibinfo {author} {\bibfnamefont {G.}~\bibnamefont
	  {Rempe}},\ }\bibfield  {title} {\bibinfo {title} {Cavity-based quantum
	  networks with single atoms and optical photons},\ }\href
	  {https://doi.org/10.1103/RevModPhys.87.1379} {\bibfield  {journal} {\bibinfo
	  {journal} {Rev.\ Mod.\ Phys.}\ }\textbf {\bibinfo {volume} {87}},\ \bibinfo
	  {pages} {1379} (\bibinfo {year} {2015})}\BibitemShut {NoStop}%
	\bibitem [{\citenamefont {Duan}\ and\ \citenamefont {Monroe}(2010)}]{duan2010}%
	  \BibitemOpen
	  \bibfield  {author} {\bibinfo {author} {\bibfnamefont {L.-M.}\ \bibnamefont
	  {Duan}}\ and\ \bibinfo {author} {\bibfnamefont {C.}~\bibnamefont {Monroe}},\
	  }\bibfield  {title} {\bibinfo {title} {{\emph{Colloquium}}: {{Quantum}}
	  networks with trapped ions},\ }\href
	  {https://doi.org/10.1103/RevModPhys.82.1209} {\bibfield  {journal} {\bibinfo
	  {journal} {Rev.\ Mod.\ Phys.}\ }\textbf {\bibinfo {volume} {82}},\ \bibinfo
	  {pages} {1209} (\bibinfo {year} {2010})}\BibitemShut {NoStop}%
	\bibitem [{\citenamefont {Lodahl}(2017)}]{lodahl2017a}%
	  \BibitemOpen
	  \bibfield  {author} {\bibinfo {author} {\bibfnamefont {P.}~\bibnamefont
	  {Lodahl}},\ }\bibfield  {title} {\bibinfo {title} {Quantum-dot based photonic
	  quantum networks},\ }\href {https://doi.org/10.1088/2058-9565/aa91bb}
	  {\bibfield  {journal} {\bibinfo  {journal} {Quantum Sci.\ Technol.}\ }\textbf
	  {\bibinfo {volume} {3}},\ \bibinfo {pages} {013001} (\bibinfo {year}
	  {2017})}\BibitemShut {NoStop}%
	\bibitem [{\citenamefont {Awschalom}\ \emph {et~al.}(2018)\citenamefont
	  {Awschalom}, \citenamefont {Hanson}, \citenamefont {Wrachtrup},\ and\
	  \citenamefont {Zhou}}]{awschalom2018}%
	  \BibitemOpen
	  \bibfield  {author} {\bibinfo {author} {\bibfnamefont {D.~D.}\ \bibnamefont
	  {Awschalom}}, \bibinfo {author} {\bibfnamefont {R.}~\bibnamefont {Hanson}},
	  \bibinfo {author} {\bibfnamefont {J.}~\bibnamefont {Wrachtrup}},\ and\
	  \bibinfo {author} {\bibfnamefont {B.~B.}\ \bibnamefont {Zhou}},\ }\bibfield
	  {title} {\bibinfo {title} {Quantum technologies with optically interfaced
	  solid-state spins},\ }\href {https://doi.org/10.1038/s41566-018-0232-2}
	  {\bibfield  {journal} {\bibinfo  {journal} {Nature Photon.}\ }\textbf
	  {\bibinfo {volume} {12}},\ \bibinfo {pages} {516} (\bibinfo {year}
	  {2018})}\BibitemShut {NoStop}%
	\bibitem [{\citenamefont {Ruf}\ \emph {et~al.}(2021)\citenamefont {Ruf},
	  \citenamefont {Wan}, \citenamefont {Choi}, \citenamefont {Englund},\ and\
	  \citenamefont {Hanson}}]{ruf2021a}%
	  \BibitemOpen
	  \bibfield  {author} {\bibinfo {author} {\bibfnamefont {M.}~\bibnamefont
	  {Ruf}}, \bibinfo {author} {\bibfnamefont {N.~H.}\ \bibnamefont {Wan}},
	  \bibinfo {author} {\bibfnamefont {H.}~\bibnamefont {Choi}}, \bibinfo {author}
	  {\bibfnamefont {D.}~\bibnamefont {Englund}},\ and\ \bibinfo {author}
	  {\bibfnamefont {R.}~\bibnamefont {Hanson}},\ }\bibfield  {title} {\bibinfo
	  {title} {Quantum networks based on color centers in diamond},\ }\href
	  {https://doi.org/10.1063/5.0056534} {\bibfield  {journal} {\bibinfo
	  {journal} {J.\ Appl.\ Phys.}\ }\textbf {\bibinfo {volume} {130}},\ \bibinfo
	  {pages} {070901} (\bibinfo {year} {2021})}\BibitemShut {NoStop}%
	\bibitem [{\citenamefont {Pompili}\ \emph {et~al.}(2021)\citenamefont
	  {Pompili}, \citenamefont {Hermans}, \citenamefont {Baier}, \citenamefont
	  {Beukers}, \citenamefont {Humphreys}, \citenamefont {Schouten}, \citenamefont
	  {Vermeulen}, \citenamefont {Tiggelman}, \citenamefont {Martins},
	  \citenamefont {Dirkse}, \citenamefont {Wehner},\ and\ \citenamefont
	  {Hanson}}]{pompili2021}%
	  \BibitemOpen
	  \bibfield  {author} {\bibinfo {author} {\bibfnamefont {M.}~\bibnamefont
	  {Pompili}}, \bibinfo {author} {\bibfnamefont {S.~L.~N.}\ \bibnamefont
	  {Hermans}}, \bibinfo {author} {\bibfnamefont {S.}~\bibnamefont {Baier}},
	  \bibinfo {author} {\bibfnamefont {H.~K.~C.}\ \bibnamefont {Beukers}},
	  \bibinfo {author} {\bibfnamefont {P.~C.}\ \bibnamefont {Humphreys}}, \bibinfo
	  {author} {\bibfnamefont {R.~N.}\ \bibnamefont {Schouten}}, \bibinfo {author}
	  {\bibfnamefont {R.~F.~L.}\ \bibnamefont {Vermeulen}}, \bibinfo {author}
	  {\bibfnamefont {M.~J.}\ \bibnamefont {Tiggelman}}, \bibinfo {author}
	  {\bibfnamefont {L.~d.~S.}\ \bibnamefont {Martins}}, \bibinfo {author}
	  {\bibfnamefont {B.}~\bibnamefont {Dirkse}}, \bibinfo {author} {\bibfnamefont
	  {S.}~\bibnamefont {Wehner}},\ and\ \bibinfo {author} {\bibfnamefont
	  {R.}~\bibnamefont {Hanson}},\ }\bibfield  {title} {\bibinfo {title}
	  {Realization of a multinode quantum network of remote solid-state qubits},\
	  }\href {https://doi.org/10.1126/science.abg1919} {\bibfield  {journal}
	  {\bibinfo  {journal} {Science}\ }\textbf {\bibinfo {volume} {372}},\ \bibinfo
	  {pages} {259} (\bibinfo {year} {2021})}\BibitemShut {NoStop}%
	\bibitem [{\citenamefont {Hermans}\ \emph {et~al.}(2022)\citenamefont
	  {Hermans}, \citenamefont {Pompili}, \citenamefont {Beukers}, \citenamefont
	  {Baier}, \citenamefont {Borregaard},\ and\ \citenamefont
	  {Hanson}}]{hermans2022a}%
	  \BibitemOpen
	  \bibfield  {author} {\bibinfo {author} {\bibfnamefont {S.~L.~N.}\
	  \bibnamefont {Hermans}}, \bibinfo {author} {\bibfnamefont {M.}~\bibnamefont
	  {Pompili}}, \bibinfo {author} {\bibfnamefont {H.~K.~C.}\ \bibnamefont
	  {Beukers}}, \bibinfo {author} {\bibfnamefont {S.}~\bibnamefont {Baier}},
	  \bibinfo {author} {\bibfnamefont {J.}~\bibnamefont {Borregaard}},\ and\
	  \bibinfo {author} {\bibfnamefont {R.}~\bibnamefont {Hanson}},\ }\bibfield
	  {title} {\bibinfo {title} {Qubit teleportation between non-neighbouring nodes
	  in a quantum network},\ }\href {https://doi.org/10.1038/s41586-022-04697-y}
	  {\bibfield  {journal} {\bibinfo  {journal} {Nature}\ }\textbf {\bibinfo
	  {volume} {605}},\ \bibinfo {pages} {663} (\bibinfo {year}
	  {2022})}\BibitemShut {NoStop}%
	\bibitem [{\citenamefont {Bhaskar}\ \emph {et~al.}(2020)\citenamefont
	  {Bhaskar}, \citenamefont {Riedinger}, \citenamefont {Machielse},
	  \citenamefont {Levonian}, \citenamefont {Nguyen}, \citenamefont {Knall},
	  \citenamefont {Park}, \citenamefont {Englund}, \citenamefont {Lon{\v c}ar},
	  \citenamefont {Sukachev},\ and\ \citenamefont {Lukin}}]{bhaskar2020}%
	  \BibitemOpen
	  \bibfield  {author} {\bibinfo {author} {\bibfnamefont {M.~K.}\ \bibnamefont
	  {Bhaskar}}, \bibinfo {author} {\bibfnamefont {R.}~\bibnamefont {Riedinger}},
	  \bibinfo {author} {\bibfnamefont {B.}~\bibnamefont {Machielse}}, \bibinfo
	  {author} {\bibfnamefont {D.~S.}\ \bibnamefont {Levonian}}, \bibinfo {author}
	  {\bibfnamefont {C.~T.}\ \bibnamefont {Nguyen}}, \bibinfo {author}
	  {\bibfnamefont {E.~N.}\ \bibnamefont {Knall}}, \bibinfo {author}
	  {\bibfnamefont {H.}~\bibnamefont {Park}}, \bibinfo {author} {\bibfnamefont
	  {D.}~\bibnamefont {Englund}}, \bibinfo {author} {\bibfnamefont
	  {M.}~\bibnamefont {Lon{\v c}ar}}, \bibinfo {author} {\bibfnamefont {D.~D.}\
	  \bibnamefont {Sukachev}},\ and\ \bibinfo {author} {\bibfnamefont {M.~D.}\
	  \bibnamefont {Lukin}},\ }\bibfield  {title} {\bibinfo {title} {Experimental
	  demonstration of memory-enhanced quantum communication},\ }\href
	  {https://doi.org/10.1038/s41586-020-2103-5} {\bibfield  {journal} {\bibinfo
	  {journal} {Nature}\ }\textbf {\bibinfo {volume} {580}},\ \bibinfo {pages}
	  {60} (\bibinfo {year} {2020})}\BibitemShut {NoStop}%
	\bibitem [{\citenamefont {Hensen}\ \emph {et~al.}(2015)\citenamefont {Hensen},
	  \citenamefont {Bernien}, \citenamefont {Dr{\'e}au}, \citenamefont {Reiserer},
	  \citenamefont {Kalb}, \citenamefont {Blok}, \citenamefont {Ruitenberg},
	  \citenamefont {Vermeulen}, \citenamefont {Schouten}, \citenamefont
	  {Abell{\'a}n}, \citenamefont {Amaya}, \citenamefont {Pruneri}, \citenamefont
	  {Mitchell}, \citenamefont {Markham}, \citenamefont {Twitchen}, \citenamefont
	  {Elkouss}, \citenamefont {Wehner}, \citenamefont {Taminiau},\ and\
	  \citenamefont {Hanson}}]{hensen2015}%
	  \BibitemOpen
	  \bibfield  {author} {\bibinfo {author} {\bibfnamefont {B.}~\bibnamefont
	  {Hensen}}, \bibinfo {author} {\bibfnamefont {H.}~\bibnamefont {Bernien}},
	  \bibinfo {author} {\bibfnamefont {A.~E.}\ \bibnamefont {Dr{\'e}au}}, \bibinfo
	  {author} {\bibfnamefont {A.}~\bibnamefont {Reiserer}}, \bibinfo {author}
	  {\bibfnamefont {N.}~\bibnamefont {Kalb}}, \bibinfo {author} {\bibfnamefont
	  {M.~S.}\ \bibnamefont {Blok}}, \bibinfo {author} {\bibfnamefont
	  {J.}~\bibnamefont {Ruitenberg}}, \bibinfo {author} {\bibfnamefont {R.~F.~L.}\
	  \bibnamefont {Vermeulen}}, \bibinfo {author} {\bibfnamefont {R.~N.}\
	  \bibnamefont {Schouten}}, \bibinfo {author} {\bibfnamefont {C.}~\bibnamefont
	  {Abell{\'a}n}}, \bibinfo {author} {\bibfnamefont {W.}~\bibnamefont {Amaya}},
	  \bibinfo {author} {\bibfnamefont {V.}~\bibnamefont {Pruneri}}, \bibinfo
	  {author} {\bibfnamefont {M.~W.}\ \bibnamefont {Mitchell}}, \bibinfo {author}
	  {\bibfnamefont {M.}~\bibnamefont {Markham}}, \bibinfo {author} {\bibfnamefont
	  {D.~J.}\ \bibnamefont {Twitchen}}, \bibinfo {author} {\bibfnamefont
	  {D.}~\bibnamefont {Elkouss}}, \bibinfo {author} {\bibfnamefont
	  {S.}~\bibnamefont {Wehner}}, \bibinfo {author} {\bibfnamefont {T.~H.}\
	  \bibnamefont {Taminiau}},\ and\ \bibinfo {author} {\bibfnamefont
	  {R.}~\bibnamefont {Hanson}},\ }\bibfield  {title} {\bibinfo {title}
	  {Loophole-free {{Bell}} inequality violation using electron spins separated
	  by 1.3 kilometres},\ }\href {https://doi.org/10.1038/nature15759} {\bibfield
	  {journal} {\bibinfo  {journal} {Nature}\ }\textbf {\bibinfo {volume} {526}},\
	  \bibinfo {pages} {682} (\bibinfo {year} {2015})}\BibitemShut {NoStop}%
	\bibitem [{\citenamefont {Askarani}\ \emph {et~al.}(2021)\citenamefont
	  {Askarani}, \citenamefont {Das}, \citenamefont {Davidson}, \citenamefont
	  {Amaral}, \citenamefont {Sinclair}, \citenamefont {Slater}, \citenamefont
	  {Marzban}, \citenamefont {Thiel}, \citenamefont {Cone}, \citenamefont
	  {Oblak},\ and\ \citenamefont {Tittel}}]{askarani2021}%
	  \BibitemOpen
	  \bibfield  {author} {\bibinfo {author} {\bibfnamefont {M.~F.}\ \bibnamefont
	  {Askarani}}, \bibinfo {author} {\bibfnamefont {A.}~\bibnamefont {Das}},
	  \bibinfo {author} {\bibfnamefont {J.~H.}\ \bibnamefont {Davidson}}, \bibinfo
	  {author} {\bibfnamefont {G.~C.}\ \bibnamefont {Amaral}}, \bibinfo {author}
	  {\bibfnamefont {N.}~\bibnamefont {Sinclair}}, \bibinfo {author}
	  {\bibfnamefont {J.~A.}\ \bibnamefont {Slater}}, \bibinfo {author}
	  {\bibfnamefont {S.}~\bibnamefont {Marzban}}, \bibinfo {author} {\bibfnamefont
	  {C.~W.}\ \bibnamefont {Thiel}}, \bibinfo {author} {\bibfnamefont {R.~L.}\
	  \bibnamefont {Cone}}, \bibinfo {author} {\bibfnamefont {D.}~\bibnamefont
	  {Oblak}},\ and\ \bibinfo {author} {\bibfnamefont {W.}~\bibnamefont
	  {Tittel}},\ }\bibfield  {title} {\bibinfo {title} {Long-{{Lived Solid-State
	  Optical Memory}} for {{High-Rate Quantum Repeaters}}},\ }\href
	  {https://doi.org/10.1103/PhysRevLett.127.220502} {\bibfield  {journal}
	  {\bibinfo  {journal} {Phys.\ Rev.\ Lett.}\ }\textbf {\bibinfo {volume}
	  {127}},\ \bibinfo {pages} {220502} (\bibinfo {year} {2021})}\BibitemShut
	  {NoStop}%
	\bibitem [{\citenamefont {Thiel}\ \emph {et~al.}(2011)\citenamefont {Thiel},
	  \citenamefont {B{\"o}ttger},\ and\ \citenamefont {Cone}}]{thiel2011}%
	  \BibitemOpen
	  \bibfield  {author} {\bibinfo {author} {\bibfnamefont {C.~W.}\ \bibnamefont
	  {Thiel}}, \bibinfo {author} {\bibfnamefont {T.}~\bibnamefont {B{\"o}ttger}},\
	  and\ \bibinfo {author} {\bibfnamefont {R.~L.}\ \bibnamefont {Cone}},\
	  }\bibfield  {title} {\bibinfo {title} {Rare-earth-doped materials for
	  applications in quantum information storage and signal processing},\ }\href
	  {https://doi.org/10.1016/j.jlumin.2010.12.015} {\bibfield  {journal}
	  {\bibinfo  {journal} {J.\ Lumin.}\ }\bibinfo {series} {Selected Papers from
	  {{DPC}}'10},\ \textbf {\bibinfo {volume} {131}},\ \bibinfo {pages} {353}
	  (\bibinfo {year} {2011})}\BibitemShut {NoStop}%
	\bibitem [{\citenamefont {Lvovsky}\ \emph {et~al.}(2009)\citenamefont
	  {Lvovsky}, \citenamefont {Sanders},\ and\ \citenamefont
	  {Tittel}}]{lvovsky2009a}%
	  \BibitemOpen
	  \bibfield  {author} {\bibinfo {author} {\bibfnamefont {A.~I.}\ \bibnamefont
	  {Lvovsky}}, \bibinfo {author} {\bibfnamefont {B.~C.}\ \bibnamefont
	  {Sanders}},\ and\ \bibinfo {author} {\bibfnamefont {W.}~\bibnamefont
	  {Tittel}},\ }\bibfield  {title} {\bibinfo {title} {Optical quantum memory},\
	  }\href {https://doi.org/10.1038/nphoton.2009.231} {\bibfield  {journal}
	  {\bibinfo  {journal} {Nature Photon.}\ }\textbf {\bibinfo {volume} {3}},\
	  \bibinfo {pages} {706} (\bibinfo {year} {2009})}\BibitemShut {NoStop}%
	\bibitem [{\citenamefont {Kraus}\ \emph {et~al.}(2006)\citenamefont {Kraus},
	  \citenamefont {Tittel}, \citenamefont {Gisin}, \citenamefont {Nilsson},
	  \citenamefont {Kr{\"o}ll},\ and\ \citenamefont {Cirac}}]{kraus2006}%
	  \BibitemOpen
	  \bibfield  {author} {\bibinfo {author} {\bibfnamefont {B.}~\bibnamefont
	  {Kraus}}, \bibinfo {author} {\bibfnamefont {W.}~\bibnamefont {Tittel}},
	  \bibinfo {author} {\bibfnamefont {N.}~\bibnamefont {Gisin}}, \bibinfo
	  {author} {\bibfnamefont {M.}~\bibnamefont {Nilsson}}, \bibinfo {author}
	  {\bibfnamefont {S.}~\bibnamefont {Kr{\"o}ll}},\ and\ \bibinfo {author}
	  {\bibfnamefont {J.~I.}\ \bibnamefont {Cirac}},\ }\bibfield  {title} {\bibinfo
	  {title} {Quantum memory for nonstationary light fields based on controlled
	  reversible inhomogeneous broadening},\ }\href
	  {https://doi.org/10.1103/PhysRevA.73.020302} {\bibfield  {journal} {\bibinfo
	  {journal} {Phys.\ Rev.\ A}\ }\textbf {\bibinfo {volume} {73}},\ \bibinfo
	  {pages} {020302} (\bibinfo {year} {2006})}\BibitemShut {NoStop}%
	\bibitem [{\citenamefont {Afzelius}\ \emph {et~al.}(2009)\citenamefont
	  {Afzelius}, \citenamefont {Simon}, \citenamefont {{de Riedmatten}},\ and\
	  \citenamefont {Gisin}}]{afzelius2009}%
	  \BibitemOpen
	  \bibfield  {author} {\bibinfo {author} {\bibfnamefont {M.}~\bibnamefont
	  {Afzelius}}, \bibinfo {author} {\bibfnamefont {C.}~\bibnamefont {Simon}},
	  \bibinfo {author} {\bibfnamefont {H.}~\bibnamefont {{de Riedmatten}}},\ and\
	  \bibinfo {author} {\bibfnamefont {N.}~\bibnamefont {Gisin}},\ }\bibfield
	  {title} {\bibinfo {title} {Multimode quantum memory based on atomic frequency
	  combs},\ }\href {https://doi.org/10.1103/PhysRevA.79.052329} {\bibfield
	  {journal} {\bibinfo  {journal} {Phys.\ Rev.\ A}\ }\textbf {\bibinfo {volume}
	  {79}},\ \bibinfo {pages} {052329} (\bibinfo {year} {2009})}\BibitemShut
	  {NoStop}%
	\bibitem [{\citenamefont {Hedges}\ \emph {et~al.}(2010)\citenamefont {Hedges},
	  \citenamefont {Longdell}, \citenamefont {Li},\ and\ \citenamefont
	  {Sellars}}]{hedges2010}%
	  \BibitemOpen
	  \bibfield  {author} {\bibinfo {author} {\bibfnamefont {M.~P.}\ \bibnamefont
	  {Hedges}}, \bibinfo {author} {\bibfnamefont {J.~J.}\ \bibnamefont
	  {Longdell}}, \bibinfo {author} {\bibfnamefont {Y.}~\bibnamefont {Li}},\ and\
	  \bibinfo {author} {\bibfnamefont {M.~J.}\ \bibnamefont {Sellars}},\
	  }\bibfield  {title} {\bibinfo {title} {Efficient quantum memory for light},\
	  }\href {https://doi.org/10.1038/nature09081} {\bibfield  {journal} {\bibinfo
	  {journal} {Nature}\ }\textbf {\bibinfo {volume} {465}},\ \bibinfo {pages}
	  {1052} (\bibinfo {year} {2010})}\BibitemShut {NoStop}%
	\bibitem [{\citenamefont {Sabooni}\ \emph {et~al.}(2013)\citenamefont
	  {Sabooni}, \citenamefont {Li}, \citenamefont {Kr{\"o}ll},\ and\ \citenamefont
	  {Rippe}}]{sabooni2013}%
	  \BibitemOpen
	  \bibfield  {author} {\bibinfo {author} {\bibfnamefont {M.}~\bibnamefont
	  {Sabooni}}, \bibinfo {author} {\bibfnamefont {Q.}~\bibnamefont {Li}},
	  \bibinfo {author} {\bibfnamefont {S.}~\bibnamefont {Kr{\"o}ll}},\ and\
	  \bibinfo {author} {\bibfnamefont {L.}~\bibnamefont {Rippe}},\ }\bibfield
	  {title} {\bibinfo {title} {Efficient {{Quantum Memory Using}} a {{Weakly
	  Absorbing Sample}}},\ }\href {https://doi.org/10.1103/PhysRevLett.110.133604}
	  {\bibfield  {journal} {\bibinfo  {journal} {Phys.\ Rev.\ Lett.}\ }\textbf
	  {\bibinfo {volume} {110}},\ \bibinfo {pages} {133604} (\bibinfo {year}
	  {2013})}\BibitemShut {NoStop}%
	\bibitem [{\citenamefont {Davidson}\ \emph {et~al.}(2020)\citenamefont
	  {Davidson}, \citenamefont {Lefebvre}, \citenamefont {Zhang}, \citenamefont
	  {Oblak},\ and\ \citenamefont {Tittel}}]{davidson2020}%
	  \BibitemOpen
	  \bibfield  {author} {\bibinfo {author} {\bibfnamefont {J.~H.}\ \bibnamefont
	  {Davidson}}, \bibinfo {author} {\bibfnamefont {P.}~\bibnamefont {Lefebvre}},
	  \bibinfo {author} {\bibfnamefont {J.}~\bibnamefont {Zhang}}, \bibinfo
	  {author} {\bibfnamefont {D.}~\bibnamefont {Oblak}},\ and\ \bibinfo {author}
	  {\bibfnamefont {W.}~\bibnamefont {Tittel}},\ }\bibfield  {title} {\bibinfo
	  {title} {Improved light-matter interaction for storage of quantum states of
	  light in a thulium-doped crystal cavity},\ }\href
	  {https://doi.org/10.1103/PhysRevA.101.042333} {\bibfield  {journal} {\bibinfo
	   {journal} {Phys.\ Rev.\ A}\ }\textbf {\bibinfo {volume} {101}},\ \bibinfo
	  {pages} {042333} (\bibinfo {year} {2020})}\BibitemShut {NoStop}%
	\bibitem [{\citenamefont {Ortu}\ \emph {et~al.}(2022)\citenamefont {Ortu},
	  \citenamefont {Holz{\"a}pfel}, \citenamefont {Etesse},\ and\ \citenamefont
	  {Afzelius}}]{ortu2022a}%
	  \BibitemOpen
	  \bibfield  {author} {\bibinfo {author} {\bibfnamefont {A.}~\bibnamefont
	  {Ortu}}, \bibinfo {author} {\bibfnamefont {A.}~\bibnamefont {Holz{\"a}pfel}},
	  \bibinfo {author} {\bibfnamefont {J.}~\bibnamefont {Etesse}},\ and\ \bibinfo
	  {author} {\bibfnamefont {M.}~\bibnamefont {Afzelius}},\ }\bibfield  {title}
	  {\bibinfo {title} {Storage of photonic time-bin qubits for up to 20 ms in a
	  rare-earth doped crystal},\ }\href
	  {https://doi.org/10.1038/s41534-022-00541-3} {\bibfield  {journal} {\bibinfo
	  {journal} {npj Quantum Inf.}\ }\textbf {\bibinfo {volume} {8}},\ \bibinfo
	  {pages} {1} (\bibinfo {year} {2022})}\BibitemShut {NoStop}%
	\bibitem [{\citenamefont {Ma}\ \emph {et~al.}(2021)\citenamefont {Ma},
	  \citenamefont {Ma}, \citenamefont {Zhou}, \citenamefont {Li},\ and\
	  \citenamefont {Guo}}]{ma2021a}%
	  \BibitemOpen
	  \bibfield  {author} {\bibinfo {author} {\bibfnamefont {Y.}~\bibnamefont
	  {Ma}}, \bibinfo {author} {\bibfnamefont {Y.-Z.}\ \bibnamefont {Ma}}, \bibinfo
	  {author} {\bibfnamefont {Z.-Q.}\ \bibnamefont {Zhou}}, \bibinfo {author}
	  {\bibfnamefont {C.-F.}\ \bibnamefont {Li}},\ and\ \bibinfo {author}
	  {\bibfnamefont {G.-C.}\ \bibnamefont {Guo}},\ }\bibfield  {title} {\bibinfo
	  {title} {One-hour coherent optical storage in an atomic frequency comb
	  memory},\ }\href {https://doi.org/10.1038/s41467-021-22706-y} {\bibfield
	  {journal} {\bibinfo  {journal} {Nature Commun.}\ }\textbf {\bibinfo {volume}
	  {12}},\ \bibinfo {pages} {2381} (\bibinfo {year} {2021})}\BibitemShut
	  {NoStop}%
	\bibitem [{\citenamefont {Saglamyurek}\ \emph {et~al.}(2011)\citenamefont
	  {Saglamyurek}, \citenamefont {Sinclair}, \citenamefont {Jin}, \citenamefont
	  {Slater}, \citenamefont {Oblak}, \citenamefont {Bussi{\`e}res}, \citenamefont
	  {George}, \citenamefont {Ricken}, \citenamefont {Sohler},\ and\ \citenamefont
	  {Tittel}}]{saglamyurek2011}%
	  \BibitemOpen
	  \bibfield  {author} {\bibinfo {author} {\bibfnamefont {E.}~\bibnamefont
	  {Saglamyurek}}, \bibinfo {author} {\bibfnamefont {N.}~\bibnamefont
	  {Sinclair}}, \bibinfo {author} {\bibfnamefont {J.}~\bibnamefont {Jin}},
	  \bibinfo {author} {\bibfnamefont {J.~A.}\ \bibnamefont {Slater}}, \bibinfo
	  {author} {\bibfnamefont {D.}~\bibnamefont {Oblak}}, \bibinfo {author}
	  {\bibfnamefont {F.}~\bibnamefont {Bussi{\`e}res}}, \bibinfo {author}
	  {\bibfnamefont {M.}~\bibnamefont {George}}, \bibinfo {author} {\bibfnamefont
	  {R.}~\bibnamefont {Ricken}}, \bibinfo {author} {\bibfnamefont
	  {W.}~\bibnamefont {Sohler}},\ and\ \bibinfo {author} {\bibfnamefont
	  {W.}~\bibnamefont {Tittel}},\ }\bibfield  {title} {\bibinfo {title}
	  {Broadband waveguide quantum memory for entangled photons},\ }\href
	  {https://doi.org/10.1038/nature09719} {\bibfield  {journal} {\bibinfo
	  {journal} {Nature}\ }\textbf {\bibinfo {volume} {469}},\ \bibinfo {pages}
	  {512} (\bibinfo {year} {2011})}\BibitemShut {NoStop}%
	\bibitem [{\citenamefont {Zhou}\ \emph {et~al.}(2012)\citenamefont {Zhou},
	  \citenamefont {Lin}, \citenamefont {Yang}, \citenamefont {Li},\ and\
	  \citenamefont {Guo}}]{zhou2012}%
	  \BibitemOpen
	  \bibfield  {author} {\bibinfo {author} {\bibfnamefont {Z.-Q.}\ \bibnamefont
	  {Zhou}}, \bibinfo {author} {\bibfnamefont {W.-B.}\ \bibnamefont {Lin}},
	  \bibinfo {author} {\bibfnamefont {M.}~\bibnamefont {Yang}}, \bibinfo {author}
	  {\bibfnamefont {C.-F.}\ \bibnamefont {Li}},\ and\ \bibinfo {author}
	  {\bibfnamefont {G.-C.}\ \bibnamefont {Guo}},\ }\bibfield  {title} {\bibinfo
	  {title} {Realization of {{Reliable Solid-State Quantum Memory}} for
	  {{Photonic Polarization Qubit}}},\ }\href
	  {https://doi.org/10.1103/PhysRevLett.108.190505} {\bibfield  {journal}
	  {\bibinfo  {journal} {Phys.\ Rev.\ Lett.}\ }\textbf {\bibinfo {volume}
	  {108}},\ \bibinfo {pages} {190505} (\bibinfo {year} {2012})}\BibitemShut
	  {NoStop}%
	\bibitem [{\citenamefont {Yang}\ \emph {et~al.}(2018)\citenamefont {Yang},
	  \citenamefont {Zhou}, \citenamefont {Hua}, \citenamefont {Liu}, \citenamefont
	  {Li}, \citenamefont {Li}, \citenamefont {Ma}, \citenamefont {Liu},
	  \citenamefont {Liang}, \citenamefont {Li}, \citenamefont {Xiao},
	  \citenamefont {Hu}, \citenamefont {Li},\ and\ \citenamefont
	  {Guo}}]{yang2018a}%
	  \BibitemOpen
	  \bibfield  {author} {\bibinfo {author} {\bibfnamefont {T.-S.}\ \bibnamefont
	  {Yang}}, \bibinfo {author} {\bibfnamefont {Z.-Q.}\ \bibnamefont {Zhou}},
	  \bibinfo {author} {\bibfnamefont {Y.-L.}\ \bibnamefont {Hua}}, \bibinfo
	  {author} {\bibfnamefont {X.}~\bibnamefont {Liu}}, \bibinfo {author}
	  {\bibfnamefont {Z.-F.}\ \bibnamefont {Li}}, \bibinfo {author} {\bibfnamefont
	  {P.-Y.}\ \bibnamefont {Li}}, \bibinfo {author} {\bibfnamefont
	  {Y.}~\bibnamefont {Ma}}, \bibinfo {author} {\bibfnamefont {C.}~\bibnamefont
	  {Liu}}, \bibinfo {author} {\bibfnamefont {P.-J.}\ \bibnamefont {Liang}},
	  \bibinfo {author} {\bibfnamefont {X.}~\bibnamefont {Li}}, \bibinfo {author}
	  {\bibfnamefont {Y.-X.}\ \bibnamefont {Xiao}}, \bibinfo {author}
	  {\bibfnamefont {J.}~\bibnamefont {Hu}}, \bibinfo {author} {\bibfnamefont
	  {C.-F.}\ \bibnamefont {Li}},\ and\ \bibinfo {author} {\bibfnamefont {G.-C.}\
	  \bibnamefont {Guo}},\ }\bibfield  {title} {\bibinfo {title} {Multiplexed
	  storage and real-time manipulation based on a multiple degree-of-freedom
	  quantum memory},\ }\href {https://doi.org/10.1038/s41467-018-05669-5}
	  {\bibfield  {journal} {\bibinfo  {journal} {Nature Commun.}\ }\textbf
	  {\bibinfo {volume} {9}},\ \bibinfo {pages} {3407} (\bibinfo {year}
	  {2018})}\BibitemShut {NoStop}%
	\bibitem [{\citenamefont {Kolesov}\ \emph {et~al.}(2012)\citenamefont
	  {Kolesov}, \citenamefont {Xia}, \citenamefont {Reuter}, \citenamefont
	  {St{\"o}hr}, \citenamefont {Zappe}, \citenamefont {Meijer}, \citenamefont
	  {Hemmer},\ and\ \citenamefont {Wrachtrup}}]{kolesov2012}%
	  \BibitemOpen
	  \bibfield  {author} {\bibinfo {author} {\bibfnamefont {R.}~\bibnamefont
	  {Kolesov}}, \bibinfo {author} {\bibfnamefont {K.}~\bibnamefont {Xia}},
	  \bibinfo {author} {\bibfnamefont {R.}~\bibnamefont {Reuter}}, \bibinfo
	  {author} {\bibfnamefont {R.}~\bibnamefont {St{\"o}hr}}, \bibinfo {author}
	  {\bibfnamefont {A.}~\bibnamefont {Zappe}}, \bibinfo {author} {\bibfnamefont
	  {J.}~\bibnamefont {Meijer}}, \bibinfo {author} {\bibfnamefont
	  {P.}~\bibnamefont {Hemmer}},\ and\ \bibinfo {author} {\bibfnamefont
	  {J.}~\bibnamefont {Wrachtrup}},\ }\bibfield  {title} {\bibinfo {title}
	  {Optical detection of a single rare-earth ion in a crystal},\ }\href
	  {https://doi.org/10.1038/ncomms2034} {\bibfield  {journal} {\bibinfo
	  {journal} {Nature Commun.}\ }\textbf {\bibinfo {volume} {3}},\ \bibinfo
	  {pages} {1029} (\bibinfo {year} {2012})}\BibitemShut {NoStop}%
	\bibitem [{\citenamefont {Dibos}\ \emph {et~al.}(2018)\citenamefont {Dibos},
	  \citenamefont {Raha}, \citenamefont {Phenicie},\ and\ \citenamefont
	  {Thompson}}]{dibos2018}%
	  \BibitemOpen
	  \bibfield  {author} {\bibinfo {author} {\bibfnamefont {A.~M.}\ \bibnamefont
	  {Dibos}}, \bibinfo {author} {\bibfnamefont {M.}~\bibnamefont {Raha}},
	  \bibinfo {author} {\bibfnamefont {C.~M.}\ \bibnamefont {Phenicie}},\ and\
	  \bibinfo {author} {\bibfnamefont {J.~D.}\ \bibnamefont {Thompson}},\
	  }\bibfield  {title} {\bibinfo {title} {Atomic {{Source}} of {{Single
	  Photons}} in the {{Telecom Band}}},\ }\href
	  {https://doi.org/10.1103/PhysRevLett.120.243601} {\bibfield  {journal}
	  {\bibinfo  {journal} {Phys.\ Rev.\ Lett.}\ }\textbf {\bibinfo {volume}
	  {120}},\ \bibinfo {pages} {243601} (\bibinfo {year} {2018})}\BibitemShut
	  {NoStop}%
	\bibitem [{\citenamefont {Zhong}\ \emph {et~al.}(2018)\citenamefont {Zhong},
	  \citenamefont {Kindem}, \citenamefont {Bartholomew}, \citenamefont {Rochman},
	  \citenamefont {Craiciu}, \citenamefont {Verma}, \citenamefont {Nam},
	  \citenamefont {Marsili}, \citenamefont {Shaw}, \citenamefont {Beyer},\ and\
	  \citenamefont {Faraon}}]{zhong2018}%
	  \BibitemOpen
	  \bibfield  {author} {\bibinfo {author} {\bibfnamefont {T.}~\bibnamefont
	  {Zhong}}, \bibinfo {author} {\bibfnamefont {J.~M.}\ \bibnamefont {Kindem}},
	  \bibinfo {author} {\bibfnamefont {J.~G.}\ \bibnamefont {Bartholomew}},
	  \bibinfo {author} {\bibfnamefont {J.}~\bibnamefont {Rochman}}, \bibinfo
	  {author} {\bibfnamefont {I.}~\bibnamefont {Craiciu}}, \bibinfo {author}
	  {\bibfnamefont {V.}~\bibnamefont {Verma}}, \bibinfo {author} {\bibfnamefont
	  {S.~W.}\ \bibnamefont {Nam}}, \bibinfo {author} {\bibfnamefont
	  {F.}~\bibnamefont {Marsili}}, \bibinfo {author} {\bibfnamefont {M.~D.}\
	  \bibnamefont {Shaw}}, \bibinfo {author} {\bibfnamefont {A.~D.}\ \bibnamefont
	  {Beyer}},\ and\ \bibinfo {author} {\bibfnamefont {A.}~\bibnamefont
	  {Faraon}},\ }\bibfield  {title} {\bibinfo {title} {Optically {{Addressing
	  Single Rare-Earth Ions}} in a {{Nanophotonic Cavity}}},\ }\href
	  {https://doi.org/10.1103/PhysRevLett.121.183603} {\bibfield  {journal}
	  {\bibinfo  {journal} {Phys.\ Rev.\ Lett.}\ }\textbf {\bibinfo {volume}
	  {121}},\ \bibinfo {pages} {183603} (\bibinfo {year} {2018})}\BibitemShut
	  {NoStop}%
	\bibitem [{\citenamefont {Kindem}\ \emph {et~al.}(2020)\citenamefont {Kindem},
	  \citenamefont {Ruskuc}, \citenamefont {Bartholomew}, \citenamefont {Rochman},
	  \citenamefont {Huan},\ and\ \citenamefont {Faraon}}]{kindem2020}%
	  \BibitemOpen
	  \bibfield  {author} {\bibinfo {author} {\bibfnamefont {J.~M.}\ \bibnamefont
	  {Kindem}}, \bibinfo {author} {\bibfnamefont {A.}~\bibnamefont {Ruskuc}},
	  \bibinfo {author} {\bibfnamefont {J.~G.}\ \bibnamefont {Bartholomew}},
	  \bibinfo {author} {\bibfnamefont {J.}~\bibnamefont {Rochman}}, \bibinfo
	  {author} {\bibfnamefont {Y.~Q.}\ \bibnamefont {Huan}},\ and\ \bibinfo
	  {author} {\bibfnamefont {A.}~\bibnamefont {Faraon}},\ }\bibfield  {title}
	  {\bibinfo {title} {Control and single-shot readout of an ion embedded in a
	  nanophotonic cavity},\ }\href {https://doi.org/10.1038/s41586-020-2160-9}
	  {\bibfield  {journal} {\bibinfo  {journal} {Nature}\ }\textbf {\bibinfo
	  {volume} {580}},\ \bibinfo {pages} {201} (\bibinfo {year}
	  {2020})}\BibitemShut {NoStop}%
	\bibitem [{\citenamefont {Raha}\ \emph {et~al.}(2020)\citenamefont {Raha},
	  \citenamefont {Chen}, \citenamefont {Phenicie}, \citenamefont {Ourari},
	  \citenamefont {Dibos},\ and\ \citenamefont {Thompson}}]{raha2020}%
	  \BibitemOpen
	  \bibfield  {author} {\bibinfo {author} {\bibfnamefont {M.}~\bibnamefont
	  {Raha}}, \bibinfo {author} {\bibfnamefont {S.}~\bibnamefont {Chen}}, \bibinfo
	  {author} {\bibfnamefont {C.~M.}\ \bibnamefont {Phenicie}}, \bibinfo {author}
	  {\bibfnamefont {S.}~\bibnamefont {Ourari}}, \bibinfo {author} {\bibfnamefont
	  {A.~M.}\ \bibnamefont {Dibos}},\ and\ \bibinfo {author} {\bibfnamefont
	  {J.~D.}\ \bibnamefont {Thompson}},\ }\bibfield  {title} {\bibinfo {title}
	  {Optical quantum nondemolition measurement of a single rare earth ion
	  qubit},\ }\href {https://doi.org/10.1038/s41467-020-15138-7} {\bibfield
	  {journal} {\bibinfo  {journal} {Nature Commun.}\ }\textbf {\bibinfo {volume}
	  {11}},\ \bibinfo {pages} {1605} (\bibinfo {year} {2020})}\BibitemShut
	  {NoStop}%
	\bibitem [{\citenamefont {Yang}\ \emph {et~al.}(2023)\citenamefont {Yang},
	  \citenamefont {Wang}, \citenamefont {Shen}, \citenamefont {Xie},\ and\
	  \citenamefont {Tang}}]{yang2023}%
	  \BibitemOpen
	  \bibfield  {author} {\bibinfo {author} {\bibfnamefont {L.}~\bibnamefont
	  {Yang}}, \bibinfo {author} {\bibfnamefont {S.}~\bibnamefont {Wang}}, \bibinfo
	  {author} {\bibfnamefont {M.}~\bibnamefont {Shen}}, \bibinfo {author}
	  {\bibfnamefont {J.}~\bibnamefont {Xie}},\ and\ \bibinfo {author}
	  {\bibfnamefont {H.~X.}\ \bibnamefont {Tang}},\ }\bibfield  {title} {\bibinfo
	  {title} {Controlling single rare earth ion emission in an electro-optical
	  nanocavity},\ }\href {https://doi.org/10.1038/s41467-023-37513-w} {\bibfield
	  {journal} {\bibinfo  {journal} {Nature Commun.}\ }\textbf {\bibinfo {volume}
	  {14}},\ \bibinfo {pages} {1718} (\bibinfo {year} {2023})}\BibitemShut
	  {NoStop}%
	\bibitem [{\citenamefont {Gritsch}\ \emph {et~al.}(2023)\citenamefont
	  {Gritsch}, \citenamefont {Ulanowski},\ and\ \citenamefont
	  {Reiserer}}]{gritsch2023}%
	  \BibitemOpen
	  \bibfield  {author} {\bibinfo {author} {\bibfnamefont {A.}~\bibnamefont
	  {Gritsch}}, \bibinfo {author} {\bibfnamefont {A.}~\bibnamefont {Ulanowski}},\
	  and\ \bibinfo {author} {\bibfnamefont {A.}~\bibnamefont {Reiserer}},\
	  }\bibfield  {title} {\bibinfo {title} {Purcell enhancement of single photon
	  emitters in silicon},\ }\href {https://doi.org/10.48550/arXiv.2301.07753}
	  {\bibfield  {journal} {\bibinfo  {journal} {arXiv:2301.07753}\ } (\bibinfo
	  {year} {2023})}\BibitemShut {NoStop}%
	\bibitem [{\citenamefont {Ourari}\ \emph {et~al.}(2023)\citenamefont {Ourari},
	  \citenamefont {Dusanowski}, \citenamefont {Horvath}, \citenamefont {Uysal},
	  \citenamefont {Phenicie}, \citenamefont {Stevenson}, \citenamefont {Raha},
	  \citenamefont {Chen}, \citenamefont {Cava}, \citenamefont {{de Leon}},\ and\
	  \citenamefont {Thompson}}]{ourari2023}%
	  \BibitemOpen
	  \bibfield  {author} {\bibinfo {author} {\bibfnamefont {S.}~\bibnamefont
	  {Ourari}}, \bibinfo {author} {\bibfnamefont {{\L}.}~\bibnamefont
	  {Dusanowski}}, \bibinfo {author} {\bibfnamefont {S.~P.}\ \bibnamefont
	  {Horvath}}, \bibinfo {author} {\bibfnamefont {M.~T.}\ \bibnamefont {Uysal}},
	  \bibinfo {author} {\bibfnamefont {C.~M.}\ \bibnamefont {Phenicie}}, \bibinfo
	  {author} {\bibfnamefont {P.}~\bibnamefont {Stevenson}}, \bibinfo {author}
	  {\bibfnamefont {M.}~\bibnamefont {Raha}}, \bibinfo {author} {\bibfnamefont
	  {S.}~\bibnamefont {Chen}}, \bibinfo {author} {\bibfnamefont {R.~J.}\
	  \bibnamefont {Cava}}, \bibinfo {author} {\bibfnamefont {N.~P.}\ \bibnamefont
	  {{de Leon}}},\ and\ \bibinfo {author} {\bibfnamefont {J.~D.}\ \bibnamefont
	  {Thompson}},\ }\bibfield  {title} {\bibinfo {title} {Indistinguishable
	  telecom band photons from a single erbium ion in the solid state},\ }\href
	  {https://doi.org/10.48550/arXiv.2301.03564} {\bibfield  {journal} {\bibinfo
	  {journal} {arXiv:2301.03564}\ } (\bibinfo {year} {2023})}\BibitemShut
	  {NoStop}%
	\bibitem [{\citenamefont {Deshmukh}\ \emph {et~al.}(2023)\citenamefont
	  {Deshmukh}, \citenamefont {Beattie}, \citenamefont {Casabone}, \citenamefont
	  {Grandi}, \citenamefont {Serrano}, \citenamefont {Ferrier}, \citenamefont
	  {Goldner}, \citenamefont {Hunger},\ and\ \citenamefont {{de
	  Riedmatten}}}]{deshmukh2023}%
	  \BibitemOpen
	  \bibfield  {author} {\bibinfo {author} {\bibfnamefont {C.}~\bibnamefont
	  {Deshmukh}}, \bibinfo {author} {\bibfnamefont {E.}~\bibnamefont {Beattie}},
	  \bibinfo {author} {\bibfnamefont {B.}~\bibnamefont {Casabone}}, \bibinfo
	  {author} {\bibfnamefont {S.}~\bibnamefont {Grandi}}, \bibinfo {author}
	  {\bibfnamefont {D.}~\bibnamefont {Serrano}}, \bibinfo {author} {\bibfnamefont
	  {A.}~\bibnamefont {Ferrier}}, \bibinfo {author} {\bibfnamefont
	  {P.}~\bibnamefont {Goldner}}, \bibinfo {author} {\bibfnamefont
	  {D.}~\bibnamefont {Hunger}},\ and\ \bibinfo {author} {\bibfnamefont
	  {H.}~\bibnamefont {{de Riedmatten}}},\ }\bibfield  {title} {\bibinfo {title}
	  {Detection of single ions in a nanoparticle coupled to a fiber cavity},\
	  }\href {https://doi.org/10.48550/arXiv.2303.00017} {\bibfield  {journal}
	  {\bibinfo  {journal} {arXiv:2303.00017}\ } (\bibinfo {year}
	  {2023})}\BibitemShut {NoStop}%
	\bibitem [{\citenamefont {Kinos}\ \emph {et~al.}(2021)\citenamefont {Kinos},
	  \citenamefont {Hunger}, \citenamefont {Kolesov}, \citenamefont {M{\o}lmer},
	  \citenamefont {{de Riedmatten}}, \citenamefont {Goldner}, \citenamefont
	  {Tallaire}, \citenamefont {Morvan}, \citenamefont {Berger}, \citenamefont
	  {Welinski}, \citenamefont {Karrai}, \citenamefont {Rippe}, \citenamefont
	  {Kr{\"o}ll},\ and\ \citenamefont {Walther}}]{kinos2021a}%
	  \BibitemOpen
	  \bibfield  {author} {\bibinfo {author} {\bibfnamefont {A.}~\bibnamefont
	  {Kinos}}, \bibinfo {author} {\bibfnamefont {D.}~\bibnamefont {Hunger}},
	  \bibinfo {author} {\bibfnamefont {R.}~\bibnamefont {Kolesov}}, \bibinfo
	  {author} {\bibfnamefont {K.}~\bibnamefont {M{\o}lmer}}, \bibinfo {author}
	  {\bibfnamefont {H.}~\bibnamefont {{de Riedmatten}}}, \bibinfo {author}
	  {\bibfnamefont {P.}~\bibnamefont {Goldner}}, \bibinfo {author} {\bibfnamefont
	  {A.}~\bibnamefont {Tallaire}}, \bibinfo {author} {\bibfnamefont
	  {L.}~\bibnamefont {Morvan}}, \bibinfo {author} {\bibfnamefont
	  {P.}~\bibnamefont {Berger}}, \bibinfo {author} {\bibfnamefont
	  {S.}~\bibnamefont {Welinski}}, \bibinfo {author} {\bibfnamefont
	  {K.}~\bibnamefont {Karrai}}, \bibinfo {author} {\bibfnamefont
	  {L.}~\bibnamefont {Rippe}}, \bibinfo {author} {\bibfnamefont
	  {S.}~\bibnamefont {Kr{\"o}ll}},\ and\ \bibinfo {author} {\bibfnamefont
	  {A.}~\bibnamefont {Walther}},\ }\bibfield  {title} {\bibinfo {title} {Roadmap
	  for {{Rare-earth Quantum Computing}}},\ }\href
	  {https://doi.org/10.48550/arXiv.2103.15743} {\bibfield  {journal} {\bibinfo
	  {journal} {arXiv:2103.15743}\ } (\bibinfo {year} {2021})}\BibitemShut
	  {NoStop}%
	\bibitem [{\citenamefont {Zhong}\ \emph {et~al.}(2015)\citenamefont {Zhong},
	  \citenamefont {Hedges}, \citenamefont {Ahlefeldt}, \citenamefont
	  {Bartholomew}, \citenamefont {Beavan}, \citenamefont {Wittig}, \citenamefont
	  {Longdell},\ and\ \citenamefont {Sellars}}]{zhong2015a}%
	  \BibitemOpen
	  \bibfield  {author} {\bibinfo {author} {\bibfnamefont {M.}~\bibnamefont
	  {Zhong}}, \bibinfo {author} {\bibfnamefont {M.~P.}\ \bibnamefont {Hedges}},
	  \bibinfo {author} {\bibfnamefont {R.~L.}\ \bibnamefont {Ahlefeldt}}, \bibinfo
	  {author} {\bibfnamefont {J.~G.}\ \bibnamefont {Bartholomew}}, \bibinfo
	  {author} {\bibfnamefont {S.~E.}\ \bibnamefont {Beavan}}, \bibinfo {author}
	  {\bibfnamefont {S.~M.}\ \bibnamefont {Wittig}}, \bibinfo {author}
	  {\bibfnamefont {J.~J.}\ \bibnamefont {Longdell}},\ and\ \bibinfo {author}
	  {\bibfnamefont {M.~J.}\ \bibnamefont {Sellars}},\ }\bibfield  {title}
	  {\bibinfo {title} {Optically addressable nuclear spins in a solid with a
	  six-hour coherence time},\ }\href {https://doi.org/10.1038/nature14025}
	  {\bibfield  {journal} {\bibinfo  {journal} {Nature}\ }\textbf {\bibinfo
	  {volume} {517}},\ \bibinfo {pages} {177} (\bibinfo {year}
	  {2015})}\BibitemShut {NoStop}%
	\bibitem [{\citenamefont {Ruskuc}\ \emph {et~al.}(2022)\citenamefont {Ruskuc},
	  \citenamefont {Wu}, \citenamefont {Rochman}, \citenamefont {Choi},\ and\
	  \citenamefont {Faraon}}]{ruskuc2022}%
	  \BibitemOpen
	  \bibfield  {author} {\bibinfo {author} {\bibfnamefont {A.}~\bibnamefont
	  {Ruskuc}}, \bibinfo {author} {\bibfnamefont {C.-J.}\ \bibnamefont {Wu}},
	  \bibinfo {author} {\bibfnamefont {J.}~\bibnamefont {Rochman}}, \bibinfo
	  {author} {\bibfnamefont {J.}~\bibnamefont {Choi}},\ and\ \bibinfo {author}
	  {\bibfnamefont {A.}~\bibnamefont {Faraon}},\ }\bibfield  {title} {\bibinfo
	  {title} {Nuclear spin-wave quantum register for a solid-state qubit},\ }\href
	  {https://doi.org/10.1038/s41586-021-04293-6} {\bibfield  {journal} {\bibinfo
	  {journal} {Nature}\ }\textbf {\bibinfo {volume} {602}},\ \bibinfo {pages}
	  {408} (\bibinfo {year} {2022})}\BibitemShut {NoStop}%
	\bibitem [{\citenamefont {Uysal}\ \emph {et~al.}(2023)\citenamefont {Uysal},
	  \citenamefont {Raha}, \citenamefont {Chen}, \citenamefont {Phenicie},
	  \citenamefont {Ourari}, \citenamefont {Wang}, \citenamefont {{Van de Walle}},
	  \citenamefont {Dobrovitski},\ and\ \citenamefont {Thompson}}]{uysal2023}%
	  \BibitemOpen
	  \bibfield  {author} {\bibinfo {author} {\bibfnamefont {M.~T.}\ \bibnamefont
	  {Uysal}}, \bibinfo {author} {\bibfnamefont {M.}~\bibnamefont {Raha}},
	  \bibinfo {author} {\bibfnamefont {S.}~\bibnamefont {Chen}}, \bibinfo {author}
	  {\bibfnamefont {C.~M.}\ \bibnamefont {Phenicie}}, \bibinfo {author}
	  {\bibfnamefont {S.}~\bibnamefont {Ourari}}, \bibinfo {author} {\bibfnamefont
	  {M.}~\bibnamefont {Wang}}, \bibinfo {author} {\bibfnamefont {C.~G.}\
	  \bibnamefont {{Van de Walle}}}, \bibinfo {author} {\bibfnamefont {V.~V.}\
	  \bibnamefont {Dobrovitski}},\ and\ \bibinfo {author} {\bibfnamefont {J.~D.}\
	  \bibnamefont {Thompson}},\ }\bibfield  {title} {\bibinfo {title} {Coherent
	  control of a nuclear spin via interactions with a rare-earth ion in the
	  solid-state},\ }\href {https://doi.org/10.1103/PRXQuantum.4.010323}
	  {\bibfield  {journal} {\bibinfo  {journal} {PRX Quantum}\ }\textbf {\bibinfo
	  {volume} {4}},\ \bibinfo {pages} {010323} (\bibinfo {year}
	  {2023})}\BibitemShut {NoStop}%
	\bibitem [{\citenamefont {Barrett}\ and\ \citenamefont
	  {Kok}(2005)}]{barrett2005}%
	  \BibitemOpen
	  \bibfield  {author} {\bibinfo {author} {\bibfnamefont {S.~D.}\ \bibnamefont
	  {Barrett}}\ and\ \bibinfo {author} {\bibfnamefont {P.}~\bibnamefont {Kok}},\
	  }\bibfield  {title} {\bibinfo {title} {Efficient high-fidelity quantum
	  computation using matter qubits and linear optics},\ }\href
	  {https://doi.org/10.1103/PhysRevA.71.060310} {\bibfield  {journal} {\bibinfo
	  {journal} {Phys.\ Rev.\ A}\ }\textbf {\bibinfo {volume} {71}},\ \bibinfo
	  {pages} {060310} (\bibinfo {year} {2005})}\BibitemShut {NoStop}%
	\bibitem [{\citenamefont {Zhao}\ \emph {et~al.}(2014)\citenamefont {Zhao},
	  \citenamefont {Zhang}, \citenamefont {Yang}, \citenamefont {Sang},
	  \citenamefont {Jiang}, \citenamefont {Bao},\ and\ \citenamefont
	  {Pan}}]{zhao2014}%
	  \BibitemOpen
	  \bibfield  {author} {\bibinfo {author} {\bibfnamefont {T.-M.}\ \bibnamefont
	  {Zhao}}, \bibinfo {author} {\bibfnamefont {H.}~\bibnamefont {Zhang}},
	  \bibinfo {author} {\bibfnamefont {J.}~\bibnamefont {Yang}}, \bibinfo {author}
	  {\bibfnamefont {Z.-R.}\ \bibnamefont {Sang}}, \bibinfo {author}
	  {\bibfnamefont {X.}~\bibnamefont {Jiang}}, \bibinfo {author} {\bibfnamefont
	  {X.-H.}\ \bibnamefont {Bao}},\ and\ \bibinfo {author} {\bibfnamefont {J.-W.}\
	  \bibnamefont {Pan}},\ }\bibfield  {title} {\bibinfo {title} {Entangling
	  {{Different-Color Photons}} via {{Time-Resolved Measurement}} and {{Active
	  Feed Forward}}},\ }\href {https://doi.org/10.1103/PhysRevLett.112.103602}
	  {\bibfield  {journal} {\bibinfo  {journal} {Phys.\ Rev.\ Lett.}\ }\textbf
	  {\bibinfo {volume} {112}},\ \bibinfo {pages} {103602} (\bibinfo {year}
	  {2014})}\BibitemShut {NoStop}%
	\bibitem [{\citenamefont {Gerry}\ and\ \citenamefont
	  {Knight}(2004)}]{gerry2004}%
	  \BibitemOpen
	  \bibfield  {author} {\bibinfo {author} {\bibfnamefont {C.}~\bibnamefont
	  {Gerry}}\ and\ \bibinfo {author} {\bibfnamefont {P.}~\bibnamefont {Knight}},\
	  }\href {https://doi.org/10.1017/CBO9780511791239} {\emph {\bibinfo {title}
	  {Introductory Quantum Optics}}}\ (\bibinfo  {publisher} {Cambridge University
	  Press},\ \bibinfo {year} {2004})\BibitemShut {NoStop}%
	\bibitem [{\citenamefont {Guo}\ and\ \citenamefont
	  {Gr{\"o}blacher}(2022)}]{guo2022b}%
	  \BibitemOpen
	  \bibfield  {author} {\bibinfo {author} {\bibfnamefont {J.}~\bibnamefont
	  {Guo}}\ and\ \bibinfo {author} {\bibfnamefont {S.}~\bibnamefont
	  {Gr{\"o}blacher}},\ }\bibfield  {title} {\bibinfo {title} {Integrated
	  optical-readout of a high-{{Q}} mechanical out-of-plane mode},\ }\href
	  {https://doi.org/10.1038/s41377-022-00966-7} {\bibfield  {journal} {\bibinfo
	  {journal} {Light Sci.\ Appl.}\ }\textbf {\bibinfo {volume} {11}},\ \bibinfo
	  {pages} {282} (\bibinfo {year} {2022})}\BibitemShut {NoStop}%
	\bibitem [{\citenamefont {Kaplyanskii}(2002)}]{kaplyanskii2002}%
	  \BibitemOpen
	  \bibfield  {author} {\bibinfo {author} {\bibfnamefont {A.~A.}\ \bibnamefont
	  {Kaplyanskii}},\ }\bibfield  {title} {\bibinfo {title} {Linear {{Stark}}
	  effect in spectroscopy and luminescence of doped inorganic insulating
	  crystals},\ }\href {https://doi.org/10.1016/S0022-2313(02)00436-2} {\bibfield
	   {journal} {\bibinfo  {journal} {J.\ Lumin.}\ }\textbf {\bibinfo {volume}
	  {100}},\ \bibinfo {pages} {21} (\bibinfo {year} {2002})}\BibitemShut
	  {NoStop}%
	\bibitem [{\citenamefont {{Hastings-Simon}}\ \emph {et~al.}(2006)\citenamefont
	  {{Hastings-Simon}}, \citenamefont {Staudt}, \citenamefont {Afzelius},
	  \citenamefont {Baldi}, \citenamefont {Jaccard}, \citenamefont {Tittel},\ and\
	  \citenamefont {Gisin}}]{hastings-simon2006}%
	  \BibitemOpen
	  \bibfield  {author} {\bibinfo {author} {\bibfnamefont {S.~R.}\ \bibnamefont
	  {{Hastings-Simon}}}, \bibinfo {author} {\bibfnamefont {M.~U.}\ \bibnamefont
	  {Staudt}}, \bibinfo {author} {\bibfnamefont {M.}~\bibnamefont {Afzelius}},
	  \bibinfo {author} {\bibfnamefont {P.}~\bibnamefont {Baldi}}, \bibinfo
	  {author} {\bibfnamefont {D.}~\bibnamefont {Jaccard}}, \bibinfo {author}
	  {\bibfnamefont {W.}~\bibnamefont {Tittel}},\ and\ \bibinfo {author}
	  {\bibfnamefont {N.}~\bibnamefont {Gisin}},\ }\bibfield  {title} {\bibinfo
	  {title} {Controlled {{Stark}} shifts in {{Er3}}+-doped crystalline and
	  amorphous waveguides for quantum state storage},\ }\href
	  {https://doi.org/10.1016/j.optcom.2006.05.003} {\bibfield  {journal}
	  {\bibinfo  {journal} {Opt.\ Commun.}\ }\textbf {\bibinfo {volume} {266}},\
	  \bibinfo {pages} {716} (\bibinfo {year} {2006})}\BibitemShut {NoStop}%
	\bibitem [{\citenamefont {McAuslan}\ \emph {et~al.}(2009)\citenamefont
	  {McAuslan}, \citenamefont {Longdell},\ and\ \citenamefont
	  {Sellars}}]{mcauslan2009}%
	  \BibitemOpen
	  \bibfield  {author} {\bibinfo {author} {\bibfnamefont {D.~L.}\ \bibnamefont
	  {McAuslan}}, \bibinfo {author} {\bibfnamefont {J.~J.}\ \bibnamefont
	  {Longdell}},\ and\ \bibinfo {author} {\bibfnamefont {M.~J.}\ \bibnamefont
	  {Sellars}},\ }\bibfield  {title} {\bibinfo {title} {Strong-coupling cavity
	  {{QED}} using rare-earth-metal-ion dopants in monolithic resonators: {{What}}
	  you can do with a weak oscillator},\ }\href
	  {https://doi.org/10.1103/PhysRevA.80.062307} {\bibfield  {journal} {\bibinfo
	  {journal} {Phys.\ Rev.\ A}\ }\textbf {\bibinfo {volume} {80}},\ \bibinfo
	  {pages} {062307} (\bibinfo {year} {2009})}\BibitemShut {NoStop}%
	\bibitem [{\citenamefont {Thiel}\ \emph {et~al.}(2010)\citenamefont {Thiel},
	  \citenamefont {Macfarlane}, \citenamefont {B{\"o}ttger}, \citenamefont {Sun},
	  \citenamefont {Cone},\ and\ \citenamefont {Babbitt}}]{thiel2010}%
	  \BibitemOpen
	  \bibfield  {author} {\bibinfo {author} {\bibfnamefont {C.}~\bibnamefont
	  {Thiel}}, \bibinfo {author} {\bibfnamefont {R.}~\bibnamefont {Macfarlane}},
	  \bibinfo {author} {\bibfnamefont {T.}~\bibnamefont {B{\"o}ttger}}, \bibinfo
	  {author} {\bibfnamefont {Y.}~\bibnamefont {Sun}}, \bibinfo {author}
	  {\bibfnamefont {R.}~\bibnamefont {Cone}},\ and\ \bibinfo {author}
	  {\bibfnamefont {W.}~\bibnamefont {Babbitt}},\ }\bibfield  {title} {\bibinfo
	  {title} {Optical decoherence and persistent spectral hole burning in
	  {{Er3}}+:{{LiNbO3}}},\ }\href {https://doi.org/10.1016/j.jlumin.2009.12.020}
	  {\bibfield  {journal} {\bibinfo  {journal} {J.\ Lumin.}\ }\textbf {\bibinfo
	  {volume} {130}},\ \bibinfo {pages} {1603} (\bibinfo {year}
	  {2010})}\BibitemShut {NoStop}%
	\bibitem [{\citenamefont {Babajanyan}(2013)}]{babajanyan2013a}%
	  \BibitemOpen
	  \bibfield  {author} {\bibinfo {author} {\bibfnamefont {V.~G.}\ \bibnamefont
	  {Babajanyan}},\ }\bibfield  {title} {\bibinfo {title} {Spectroscopic study of
	  the expected optical cooling effect of {{LiNbO3}}:{{Er3}}+ crystal},\ }\href
	  {https://doi.org/10.1088/1054-660X/23/12/126002} {\bibfield  {journal}
	  {\bibinfo  {journal} {Laser Phys.}\ }\textbf {\bibinfo {volume} {23}},\
	  \bibinfo {pages} {126002} (\bibinfo {year} {2013})}\BibitemShut {NoStop}%
	\bibitem [{\citenamefont {Zhong}\ \emph {et~al.}(2017)\citenamefont {Zhong},
	  \citenamefont {Kindem}, \citenamefont {Bartholomew}, \citenamefont {Rochman},
	  \citenamefont {Craiciu}, \citenamefont {Miyazono}, \citenamefont
	  {Bettinelli}, \citenamefont {Cavalli}, \citenamefont {Verma}, \citenamefont
	  {Nam}, \citenamefont {Marsili}, \citenamefont {Shaw}, \citenamefont {Beyer},\
	  and\ \citenamefont {Faraon}}]{zhong2017a}%
	  \BibitemOpen
	  \bibfield  {author} {\bibinfo {author} {\bibfnamefont {T.}~\bibnamefont
	  {Zhong}}, \bibinfo {author} {\bibfnamefont {J.~M.}\ \bibnamefont {Kindem}},
	  \bibinfo {author} {\bibfnamefont {J.~G.}\ \bibnamefont {Bartholomew}},
	  \bibinfo {author} {\bibfnamefont {J.}~\bibnamefont {Rochman}}, \bibinfo
	  {author} {\bibfnamefont {I.}~\bibnamefont {Craiciu}}, \bibinfo {author}
	  {\bibfnamefont {E.}~\bibnamefont {Miyazono}}, \bibinfo {author}
	  {\bibfnamefont {M.}~\bibnamefont {Bettinelli}}, \bibinfo {author}
	  {\bibfnamefont {E.}~\bibnamefont {Cavalli}}, \bibinfo {author} {\bibfnamefont
	  {V.}~\bibnamefont {Verma}}, \bibinfo {author} {\bibfnamefont {S.~W.}\
	  \bibnamefont {Nam}}, \bibinfo {author} {\bibfnamefont {F.}~\bibnamefont
	  {Marsili}}, \bibinfo {author} {\bibfnamefont {M.~D.}\ \bibnamefont {Shaw}},
	  \bibinfo {author} {\bibfnamefont {A.~D.}\ \bibnamefont {Beyer}},\ and\
	  \bibinfo {author} {\bibfnamefont {A.}~\bibnamefont {Faraon}},\ }\bibfield
	  {title} {\bibinfo {title} {Nanophotonic rare-earth quantum memory with
	  optically controlled retrieval},\ }\href
	  {https://doi.org/10.1126/science.aan5959} {\bibfield  {journal} {\bibinfo
	  {journal} {Science}\ }\textbf {\bibinfo {volume} {357}},\ \bibinfo {pages}
	  {1392} (\bibinfo {year} {2017})}\BibitemShut {NoStop}%
	\bibitem [{\citenamefont {Wang}\ \emph {et~al.}(2022)\citenamefont {Wang},
	  \citenamefont {Yang}, \citenamefont {Shen}, \citenamefont {Fu}, \citenamefont
	  {Xu}, \citenamefont {Cone}, \citenamefont {Thiel},\ and\ \citenamefont
	  {Tang}}]{wang2022j}%
	  \BibitemOpen
	  \bibfield  {author} {\bibinfo {author} {\bibfnamefont {S.}~\bibnamefont
	  {Wang}}, \bibinfo {author} {\bibfnamefont {L.}~\bibnamefont {Yang}}, \bibinfo
	  {author} {\bibfnamefont {M.}~\bibnamefont {Shen}}, \bibinfo {author}
	  {\bibfnamefont {W.}~\bibnamefont {Fu}}, \bibinfo {author} {\bibfnamefont
	  {Y.}~\bibnamefont {Xu}}, \bibinfo {author} {\bibfnamefont {R.~L.}\
	  \bibnamefont {Cone}}, \bibinfo {author} {\bibfnamefont {C.~W.}\ \bibnamefont
	  {Thiel}},\ and\ \bibinfo {author} {\bibfnamefont {H.~X.}\ \bibnamefont
	  {Tang}},\ }\bibfield  {title} {\bibinfo {title} {Er:{{LiNbO}}\_3 with {{High
	  Optical Coherence Enabling Optical Thickness Control}}},\ }\href
	  {https://doi.org/10.1103/PhysRevApplied.18.014069} {\bibfield  {journal}
	  {\bibinfo  {journal} {Phys.\ Rev.\ Applied}\ }\textbf {\bibinfo {volume}
	  {18}},\ \bibinfo {pages} {014069} (\bibinfo {year} {2022})}\BibitemShut
	  {NoStop}%
	\bibitem [{\citenamefont {Patel}\ \emph {et~al.}(2010)\citenamefont {Patel},
	  \citenamefont {Bennett}, \citenamefont {Farrer}, \citenamefont {Nicoll},
	  \citenamefont {Ritchie},\ and\ \citenamefont {Shields}}]{patel2010}%
	  \BibitemOpen
	  \bibfield  {author} {\bibinfo {author} {\bibfnamefont {R.~B.}\ \bibnamefont
	  {Patel}}, \bibinfo {author} {\bibfnamefont {A.~J.}\ \bibnamefont {Bennett}},
	  \bibinfo {author} {\bibfnamefont {I.}~\bibnamefont {Farrer}}, \bibinfo
	  {author} {\bibfnamefont {C.~A.}\ \bibnamefont {Nicoll}}, \bibinfo {author}
	  {\bibfnamefont {D.~A.}\ \bibnamefont {Ritchie}},\ and\ \bibinfo {author}
	  {\bibfnamefont {A.~J.}\ \bibnamefont {Shields}},\ }\bibfield  {title}
	  {\bibinfo {title} {Two-photon interference of the emission from electrically
	  tunable remote quantum dots},\ }\href
	  {https://doi.org/10.1038/nphoton.2010.161} {\bibfield  {journal} {\bibinfo
	  {journal} {Nature Photon.}\ }\textbf {\bibinfo {volume} {4}},\ \bibinfo
	  {pages} {632} (\bibinfo {year} {2010})}\BibitemShut {NoStop}%
	\bibitem [{\citenamefont {Flagg}\ \emph {et~al.}(2010)\citenamefont {Flagg},
	  \citenamefont {Muller}, \citenamefont {Polyakov}, \citenamefont {Ling},
	  \citenamefont {Migdall},\ and\ \citenamefont {Solomon}}]{flagg2010}%
	  \BibitemOpen
	  \bibfield  {author} {\bibinfo {author} {\bibfnamefont {E.~B.}\ \bibnamefont
	  {Flagg}}, \bibinfo {author} {\bibfnamefont {A.}~\bibnamefont {Muller}},
	  \bibinfo {author} {\bibfnamefont {S.~V.}\ \bibnamefont {Polyakov}}, \bibinfo
	  {author} {\bibfnamefont {A.}~\bibnamefont {Ling}}, \bibinfo {author}
	  {\bibfnamefont {A.}~\bibnamefont {Migdall}},\ and\ \bibinfo {author}
	  {\bibfnamefont {G.~S.}\ \bibnamefont {Solomon}},\ }\bibfield  {title}
	  {\bibinfo {title} {Interference of {{Single Photons}} from {{Two Separate
	  Semiconductor Quantum Dots}}},\ }\href
	  {https://doi.org/10.1103/PhysRevLett.104.137401} {\bibfield  {journal}
	  {\bibinfo  {journal} {Phys.\ Rev.\ Lett.}\ }\textbf {\bibinfo {volume}
	  {104}},\ \bibinfo {pages} {137401} (\bibinfo {year} {2010})}\BibitemShut
	  {NoStop}%
	\end{thebibliography}

\begin{thebibliography}{7}%
		\makeatletter
		\providecommand \@ifxundefined [1]{%
		 \@ifx{#1\undefined}
		}%
		\providecommand \@ifnum [1]{%
		 \ifnum #1\expandafter \@firstoftwo
		 \else \expandafter \@secondoftwo
		 \fi
		}%
		\providecommand \@ifx [1]{%
		 \ifx #1\expandafter \@firstoftwo
		 \else \expandafter \@secondoftwo
		 \fi
		}%
		\providecommand \natexlab [1]{#1}%
		\providecommand \enquote  [1]{``#1''}%
		\providecommand \bibnamefont  [1]{#1}%
		\providecommand \bibfnamefont [1]{#1}%
		\providecommand \citenamefont [1]{#1}%
		\providecommand \href@noop [0]{\@secondoftwo}%
		\providecommand \href [0]{\begingroup \@sanitize@url \@href}%
		\providecommand \@href[1]{\@@startlink{#1}\@@href}%
		\providecommand \@@href[1]{\endgroup#1\@@endlink}%
		\providecommand \@sanitize@url [0]{\catcode `\\12\catcode `\$12\catcode
		  `\&12\catcode `\#12\catcode `\^12\catcode `\_12\catcode `\%12\relax}%
		\providecommand \@@startlink[1]{}%
		\providecommand \@@endlink[0]{}%
		\providecommand \url  [0]{\begingroup\@sanitize@url \@url }%
		\providecommand \@url [1]{\endgroup\@href {#1}{\urlprefix }}%
		\providecommand \urlprefix  [0]{URL }%
		\providecommand \Eprint [0]{\href }%
		\providecommand \doibase [0]{https://doi.org/}%
		\providecommand \selectlanguage [0]{\@gobble}%
		\providecommand \bibinfo  [0]{\@secondoftwo}%
		\providecommand \bibfield  [0]{\@secondoftwo}%
		\providecommand \translation [1]{[#1]}%
		\providecommand \BibitemOpen [0]{}%
		\providecommand \bibitemStop [0]{}%
		\providecommand \bibitemNoStop [0]{.\EOS\space}%
		\providecommand \EOS [0]{\spacefactor3000\relax}%
		\providecommand \BibitemShut  [1]{\csname bibitem#1\endcsname}%
		\let\auto@bib@innerbib\@empty
		\bibitem [{\citenamefont {Guo}\ and\ \citenamefont
		  {Gr{\"o}blacher}(2022)}]{guo2022b_sm}%
		  \BibitemOpen
		  \bibfield  {author} {\bibinfo {author} {\bibfnamefont {J.}~\bibnamefont
		  {Guo}}\ and\ \bibinfo {author} {\bibfnamefont {S.}~\bibnamefont
		  {Gr{\"o}blacher}},\ }\bibfield  {title} {\bibinfo {title} {Integrated
		  optical-readout of a high-{{Q}} mechanical out-of-plane mode},\ }\href
		  {https://doi.org/10.1038/s41377-022-00966-7} {\bibfield  {journal} {\bibinfo
		  {journal} {Light Sci.\ Appl.}\ }\textbf {\bibinfo {volume} {11}},\ \bibinfo
		  {pages} {282} (\bibinfo {year} {2022})}\BibitemShut {NoStop}%
		\bibitem [{\citenamefont {Kaplyanskii}\ and\ \citenamefont
		  {Macfarlane}(1987)}]{kaplianskii1987}%
		  \BibitemOpen
		  \bibinfo {editor} {\bibfnamefont {A.~A.}\ \bibnamefont {Kaplyanskii}}\ and\
		  \bibinfo {editor} {\bibfnamefont {R.~M.}\ \bibnamefont {Macfarlane}},\ eds.,\
		  \href@noop {} {\emph {\bibinfo {title} {Spectroscopy of Solids Containing
		  Rare Earth Ions}}}\ (\bibinfo  {publisher} {Elsevier},\ \bibinfo {year}
		  {1987})\BibitemShut {NoStop}%
		\bibitem [{\citenamefont {Macfarlane}(2002)}]{macfarlane2002}%
		  \BibitemOpen
		  \bibfield  {author} {\bibinfo {author} {\bibfnamefont {R.~M.}\ \bibnamefont
		  {Macfarlane}},\ }\bibfield  {title} {\bibinfo {title} {High-resolution laser
		  spectroscopy of rare-earth doped insulators: a personal perspective},\ }\href
		  {https://doi.org/10.1016/S0022-2313(02)00450-7} {\bibfield  {journal}
		  {\bibinfo  {journal} {J.\ Lumin.}\ }\textbf {\bibinfo {volume} {100}},\
		  \bibinfo {pages} {1} (\bibinfo {year} {2002})}\BibitemShut {NoStop}%
		\bibitem [{\citenamefont {Falamarzi~Askarani}\ \emph
		  {et~al.}(2019)\citenamefont {Falamarzi~Askarani}, \citenamefont {Puigibert},
		  \citenamefont {Lutz}, \citenamefont {Verma}, \citenamefont {Shaw},
		  \citenamefont {Nam}, \citenamefont {Sinclair}, \citenamefont {Oblak},\ and\
		  \citenamefont {Tittel}}]{askarani2019}%
		  \BibitemOpen
		  \bibfield  {author} {\bibinfo {author} {\bibfnamefont {M.}~\bibnamefont
		  {Falamarzi~Askarani}}, \bibinfo {author} {\bibfnamefont {M.~G.}\ \bibnamefont
		  {Puigibert}}, \bibinfo {author} {\bibfnamefont {T.}~\bibnamefont {Lutz}},
		  \bibinfo {author} {\bibfnamefont {V.~B.}\ \bibnamefont {Verma}}, \bibinfo
		  {author} {\bibfnamefont {M.~D.}\ \bibnamefont {Shaw}}, \bibinfo {author}
		  {\bibfnamefont {S.~W.}\ \bibnamefont {Nam}}, \bibinfo {author} {\bibfnamefont
		  {N.}~\bibnamefont {Sinclair}}, \bibinfo {author} {\bibfnamefont
		  {D.}~\bibnamefont {Oblak}},\ and\ \bibinfo {author} {\bibfnamefont
		  {W.}~\bibnamefont {Tittel}},\ }\bibfield  {title} {\bibinfo {title} {Storage
		  and reemission of heralded telecommunication-wavelength photons using a
		  crystal waveguide},\ }\href
		  {https://doi.org/10.1103/PhysRevApplied.11.054056} {\bibfield  {journal}
		  {\bibinfo  {journal} {Phys.\ Rev.\ Appl.}\ }\textbf {\bibinfo {volume}
		  {11}},\ \bibinfo {pages} {054056} (\bibinfo {year} {2019})}\BibitemShut
		  {NoStop}%
		\bibitem [{\citenamefont {Lethokov}(1976)}]{lethokov1976}%
		  \BibitemOpen
		  \bibfield  {author} {\bibinfo {author} {\bibfnamefont {V.~S.}\ \bibnamefont
		  {Lethokov}},\ }\href {https://doi.org/10.1007/3-540-07719-7} {\emph {\bibinfo
		  {title} {High-Resolution Laser Spectroscopy}}},\ Topics in Applied Physics\
		  (\bibinfo  {publisher} {Springer Berlin},\ \bibinfo {year}
		  {1976})\BibitemShut {NoStop}%
		\bibitem [{\citenamefont {Thiel}\ \emph {et~al.}(2010)\citenamefont {Thiel},
		  \citenamefont {Macfarlane}, \citenamefont {B{\"o}ttger}, \citenamefont {Sun},
		  \citenamefont {Cone},\ and\ \citenamefont {Babbitt}}]{thiel2010_sm}%
		  \BibitemOpen
		  \bibfield  {author} {\bibinfo {author} {\bibfnamefont {C.}~\bibnamefont
		  {Thiel}}, \bibinfo {author} {\bibfnamefont {R.}~\bibnamefont {Macfarlane}},
		  \bibinfo {author} {\bibfnamefont {T.}~\bibnamefont {B{\"o}ttger}}, \bibinfo
		  {author} {\bibfnamefont {Y.}~\bibnamefont {Sun}}, \bibinfo {author}
		  {\bibfnamefont {R.}~\bibnamefont {Cone}},\ and\ \bibinfo {author}
		  {\bibfnamefont {W.}~\bibnamefont {Babbitt}},\ }\bibfield  {title} {\bibinfo
		  {title} {Optical decoherence and persistent spectral hole burning in
		  {{Er3}}+:{{LiNbO3}}},\ }\href {https://doi.org/10.1016/j.jlumin.2009.12.020}
		  {\bibfield  {journal} {\bibinfo  {journal} {J.\ Lumin.}\ }\textbf {\bibinfo
		  {volume} {130}},\ \bibinfo {pages} {1603} (\bibinfo {year}
		  {2010})}\BibitemShut {NoStop}%
		\bibitem [{\citenamefont {Wang}\ \emph {et~al.}(2022)\citenamefont {Wang},
		  \citenamefont {Yang}, \citenamefont {Shen}, \citenamefont {Fu}, \citenamefont
		  {Xu}, \citenamefont {Cone}, \citenamefont {Thiel},\ and\ \citenamefont
		  {Tang}}]{Wang2022}%
		  \BibitemOpen
		  \bibfield  {author} {\bibinfo {author} {\bibfnamefont {S.}~\bibnamefont
		  {Wang}}, \bibinfo {author} {\bibfnamefont {L.}~\bibnamefont {Yang}}, \bibinfo
		  {author} {\bibfnamefont {M.}~\bibnamefont {Shen}}, \bibinfo {author}
		  {\bibfnamefont {W.}~\bibnamefont {Fu}}, \bibinfo {author} {\bibfnamefont
		  {Y.}~\bibnamefont {Xu}}, \bibinfo {author} {\bibfnamefont {R.~L.}\
		  \bibnamefont {Cone}}, \bibinfo {author} {\bibfnamefont {C.~W.}\ \bibnamefont
		  {Thiel}},\ and\ \bibinfo {author} {\bibfnamefont {H.~X.}\ \bibnamefont
		  {Tang}},\ }\bibfield  {title} {\bibinfo {title}
		  {$\mathrm{Er}:{\mathrm{li}\mathrm{nb}\mathrm{o}}_{3}$ with high optical
		  coherence enabling optical thickness control},\ }\href
		  {https://doi.org/10.1103/PhysRevApplied.18.014069} {\bibfield  {journal}
		  {\bibinfo  {journal} {Phys.\ Rev.\ Applied}\ }\textbf {\bibinfo {volume}
		  {18}},\ \bibinfo {pages} {014069} (\bibinfo {year} {2022})}\BibitemShut
		  {NoStop}%
		\end{thebibliography}
\end{document}